\documentclass[%
 preprint,
 amsmath,amssymb,
aps,
]{revtex4-1}

\usepackage{graphicx}% Include figure files
\usepackage{dcolumn}% Align table columns on decimal point
\usepackage{bm}% bold math

\usepackage{color}
\usepackage{float}
\usepackage[letterpaper, portrait, margin=1in]{geometry}
\usepackage[utf8]{inputenc}
\usepackage{graphicx}
\usepackage{subcaption}
\usepackage{multirow}
\usepackage{bibunits}
\usepackage{bigints}
\graphicspath{{Images_new/}}
\usepackage{placeins}
\usepackage{afterpage}

\usepackage[pdftex,breaklinks,colorlinks,
citecolor=blue,
urlcolor=blue,
pdfsubject={LaTeX}]{hyperref}
\begin{document}

\preprint{APS/123-QED}

% \title{Grain Boundary Segregation in Hetero-phase Interfaces: Chemical Considerations}% Force line breaks with \\
\title{A phase-field approach for modeling equilibrium solute segregation at the interphase boundary in binary alloys}

\author{Sourabh B Kadambi}
\affiliation{%
 Department of Materials Science and Engineering \\
 North Carolina State University
}%
\author{Fadi Abdeljawad}
\affiliation{%
 Department of Mechanical Engineering \\
 Clemson University, Clemson, South Carolina 29634
} 
\author{Srikanth Patala}%
 \email{spatala@ncsu.edu}
\affiliation{%
 Department of Materials Science and Engineering \\
 North Carolina State University
}%

\begin{abstract}
        A number of experimental and theoretical findings in age hardening alloys suggest that specific solute  elements preferentially segregate to and reduce the energy of the interphase boundary (IB). This segregation mechanism can stabilize the precipitation microstructure against coarsening, allowing higher operating temperatures in structural applications. Herein, we present a phase field model of solute segregation to IBs that separate matrix and precipitate phases in binary alloys. The proposed modeling framework is capable of capturing bulk thermodynamics and interfacial free energies, while also accounting for various mass transport mechanisms. Analytical equilibrium solutions of one-dimensional systems are presented, and excess IB quantities are evaluated independent of the Gibbs dividing surface convention.
        With the aid of the parallel tangent construction, IB segregation isotherms are established in terms of the alloy composition and the model parameters describing the free energy functions. Under the regular solution approximation, computational studies elucidating the dependence of the IB energy and segregation levels on temperature and free energy model parameters are presented. We show that the model is consistent with the Gibbs adsorption equation; therefore, it is possible to compare the adsorption behavior predicted by the model parameters with experiments and atomistic simulations. Future work on extending the model to ternary alloys, and incorporating the effect of elastic interactions on IB segregation is expected. 

\begin{description}
\item[Keywords] Solute Segregation, Interphase Boundary, Phase-field Model, Gibbs Adsorption

\end{description}
\end{abstract}

\pacs{Valid PACS appear here}% PACS, the Physics and Astronomy
\maketitle

%\tableofcontents

% \begin{bibunit}[apsrev4-1]
\section{Introduction}
\label{sec:Intro}

A dispersion of secondary phase precipitates in a matrix has been traditionally employed to design metallic alloys with superior mechanical properties \cite{nie2012precipitation,gladman1999precipitation,martin2012precipitation}. On exposure to elevated temperatures, however, the precipitates coarsen to reduce the excess energy associated with the presence of the interphase boundary (IB) \cite{ratke2013growth,raabe2014grain}. The consequent increase in the mean precipitate size results in a deterioration in the mechanical properties \cite{gleiter2000nanostructured,gladman1999precipitation}. Coarsening, therefore, imposes a limitation on the elevated temperature applicability of precipitation-hardened alloys \cite{vaithyanathan2004multiscale,yang2016influence,gibson2010effect,jokisaari2017predicting}.  
The driving force for coarsening is the reduction in the total interfacial energy of the microstructure, which entails the total interfacial area and the specific interfacial energy $\gamma$. The IB is therefore a critical component of the microstructure that governs high-temperature stability.

Whereas the system reduces its overall interfacial area during coarsening to minimize its energy, the IB itself could be engineered to possess lower specific energy $\gamma$. Such a system is expected to be inherently more stable due to the lower driving force for coarsening. Indeed, experimental observations \cite{marquis2003mg,marquis2005coarsening,liu2016interphase,liu2017zn,kang2014determination} have shown that the stability of precipitation hardening alloys can be enhanced by microalloying with specific solute elements that segregate to the IB to reduce the interfacial energy $\gamma$.
Additionally, first-principles calculations show that certain solutes are energetically preferred at IB sites, and are thermodynamically favored to segregate \cite{marquis2006composition,amouyal2008segregation,biswas2010simultaneous,biswas2011precipitates, shin2017solute}.

These findings demonstrate the potential for engineering IBs through solute segregation. 
Select examples of segregation in multicomponent alloys are: Mg, Si, Mg+Ag, Mn+Zr at $\alpha$-Al/$\theta^\prime$-Al$_2$Cu; Mg at Al/Al$_3$Sc; Zn at Mg/Mg$_2$Sn. IB segregation has also been observed in the binary Al-Ag alloy where Ag segregation of different amounts has been observed at Al/$\gamma^\prime$-AlAg \cite{osamura1986ap,rosalie2014chemical,zhang2017bi}. While the effect of the solute segregation in the binary system on the microstructural stability is not well understood, it is of similar interest to multicomponent alloys since (i) the segregation is suggested by first principles to be of thermodynamic origin \cite{rosalie2014chemical} and (ii) such segregation resembles in structure to that found in multicomponent systems \cite{zhang2017bi}.

With input from experimental and atomistic studies, mesoscale phase-field models allow simulation of complex microstructures and their time-dependent behavior at diffusive length and time scales \cite{vaithyanathan2002multiscale,vaithyanathan2004multiscale}.
The phase-field method is flexible in allowing a variety of physical effects, especially pertaining to the simulation of the IB \cite{thornton2003modelling}. Phase-field techniques have been used to simulate microstructures of important precipitation hardening alloys by incorporating anisotropic effects of interfacial energy \cite{vaithyanathan2004multiscale}, lattice misfit and elasticity \cite{kim2017first}, and also inhomogeneous elasticity between the matrix and the precipitate \cite{kim2017first,jokisaari2017predicting}.
However, solute segregation is expected to alter IB properties (viz. IB energy and anisotropy \cite{wynblatt2006anisotropy}, lattice misfit \cite{kang2014determination}), and thereby the microstructural features (precipitate size, morphology and distribution) and evolution kinetics (growth and coarsening). Therefore, an approach to explicitly incorporate solute segregation and its effects on the IB is essential. 

In this paper, we develop a phase-field formulation to model IB segregation and its effect on IB energy, with the aim of capturing the chemical thermodynamic aspects of segregation to the IB between a solid-solution matrix and an intermetallic precipitate. First, in Section \ref{sec:background}, we present the relevant background on available diffuse-interface approaches for modeling two-phase systems and IB segregation. In Section \ref{sec:theoretical}, we present the phase-field formulation and the analytic steady-state relations for arbitrary free energies. In Section \ref{sec:regular_soln}, we assume the regular solution behavior and perform a parametric study of the steady-state solutions and demonstrate the relationship between IB parameters and temperature on IB segregation and IB energy. We also demonstrate that the model is consistent with classical interface segregation models and that it satisfies the Gibbs adsorption relation. In Section \ref{sec:discussion}, we discuss the theoretical aspects of the model and identify potential future developments using the formulation. Relevant derivations are presented in the Appendix section.
\section{Background} \label{sec:background}
 
% \textcolor{red}{Need some re-wording, especially the statement about the WBM.}
Phase-field descriptions for polymorphous two-phase systems are generally based on the models developed either by Wheeler et al. (WBM) \cite{wheeler1992phase} or Kim et al. (KKS) \cite{kim1999phase}. The IB is a diffuse region with a gradient in the properties between the bulk matrix and precipitate phases. Any point along the IB is a thermodynamic mixture of hypothetical matrix and precipitate phases.
In the WBM approach, the hypothetical phases have equivalent concentration, which is the local IB concentration. In the KKS approach, the hypothetical phases have distinct concentrations that are constrained by equality of the diffusion potential; the local IB concentration is given by a mixture rule between the local phase concentrations. In both WBM and KKS, the local free energy density at the IB is given by a mixture rule between the local phase free energies densities. The distinct thermodynamic description of the IB in the WBM (the equal concentration condition) and KKS (the equal diffusion potential condition) models results in distinct concentration and free energy dependence across the IB as discussed in Refs. \cite{kim1999phase,kim2004phase}. At steady state, the WBM method produces an inherent contribution to the IB energy $\gamma$ and solute excess, whereas the KKS method yields zero inherent contribution to $\gamma$ and solute excess. This is because the IB region in KKS is effectively a mixture of the matrix and precipitate phases at their \emph{equilibrium} bulk phase concentrations, which is not the case in WBM.

In both approaches, an external (barrier) potential is added to the model to fit the required interfacial energy, $\gamma$. Additionally, the solute excess in the WBM approach can be tuned by employing an external concentration-dependent barrier potential \cite{wheeler1992phase,mcfadden2002gibbs,umantsev2001continuum}. The variation in solute excess and its effect on $\gamma$ has been considered for a solid-liquid binary component system described using regular solution thermodynamics \cite{wheeler1992phase,mcfadden2002gibbs}. Umantsev \cite{umantsev2001continuum,umantsev2001coherency} formulated a general theoretical description of IB segregation in multicomponent alloys. Recently, segregation of Mn+Zr to Al/$\theta'$-Al$_2$Cu IB has been simulated \cite{shower2019mechanisms,shower2019temperature}. However, there are challenges in adapting the WBM approach to model segregation to matrix-precipitate systems. For instance, realistic free energies for intermetallic precipitates based on CALPHAD have a large curvature that penalize the deviation in concentration from stoichiometry. This may result in nonphysical IB properties such as very large $\gamma$ \cite{hu2007thermodynamic} (to avoid this, models based on WBM usually employ polynomial free energy functions). The KKS approach overcomes these limitations: (i) the bulk phase free energies do not contribute directly towards $\gamma$ or interface properties, and therefore thermodynamic free energy functions can be readily employed; (ii) analytic solutions for the steady-state can be obtained. The KKS approach is therefore chosen for the present work. However, IB segregation or solute excess is zero in the traditional KKS formulation. Thus, capturing IB segregation or non-zero solute excess is an objective of the present work.

% \textcolor{red}{This may result in nonphysical IB properties such as very large $\gamma$ \cite{hu2007thermodynamic}. To avoid this, models based on WBM usually employ polynomial free energy functions, which result in physically acceptable $\gamma$, but make it difficult to obtain analytic solutions at steady state \cite{provatas2011phase}. The KKS approach overcomes these limitations: (i) the bulk phase free energies do not contribute towards $\gamma$; (ii) analytic solutions for the steady-state can be obtained.} 

In isomorphous two-phase systems, interface segregation has been modeled by employing the Cahn-Hilliard formulation \cite{dregia1991equilibrium,huang1999interfacial}. Here the ternary component $C$ segregates at the IB between the $A$-rich and $B$-rich regions of a phase-separating $A$--$B$ solution. Such an approach is applicable to systems described by a single free energy with a miscibility gap in the $A$--$B$ phase diagram and favorable $A$--$C$ and $B$--$C$ interactions; here information from bulk thermodynamics suffices to describe segregation. However, an approach to model IB segregation in \emph{polymorphous} two-phase systems, and with the \emph{flexibility to define a distinct free energy-concentration dependence for the IB} is needed. Phase-field approaches that describe solute segregation at grain boundaries (GBs) effectively assign a distinct free energy dependence to the GB \cite{abdeljawad2015stabilization, abdeljawad2017grain, kim2016GBsegregation}. In these models, the GB can be regarded as an interface phase in the sense that a fundamental thermodynamic equation governs the behavior of the boundary \cite{kaplan2013review,frolov2015phases}. This allows properties of the phase-field models to be described in relation to traditional statistical thermodynamic treatments of solute segregation to interfaces \cite{kim2016GBsegregation}. Such approaches are consistent with classical segregation isotherms and satisfy Gibbs adsorption \cite{abdeljawad2015stabilization,abdeljawad2017grain}. In the context of IBs, however, such descriptions have not been formulated to the best of our knowledge. Simple non-gradient equilibrium models describing characteristic IB thermodynamics have been proposed \cite{howe1997interfaces,kadambi2017thermodynamic}. Available phase-field models for polymorphous systems (based on WBM or KKS) do not allow a distinct free energy-concentration dependence for the IB.

Following the above discussion, in the present work we formulate a phase-field model that incorporates IB segregation within the KKS (equal diffusion potential condition) framework. The model will allow the use of realistic free energies for the intermetallic precipitate phase and a distinct free energy for the IB phase. Analytic solutions for $\gamma$ and solute excess will be derived for a one-dimensional system at steady state, and the solutions will be shown to be consistent with classical segregation isotherms and the Gibbs adsorption relation.

\section{Phase-field model} \label{sec:theoretical}

\subsection{Model formulation} \label{subsec:pf_model}

\begin{figure}[h!]
% \captionsetup[subfigure]{justification=centering}
    \centering
    \begin{subfigure}[t]{0.45\textwidth}
        \centering
        \includegraphics[width=1\textwidth]{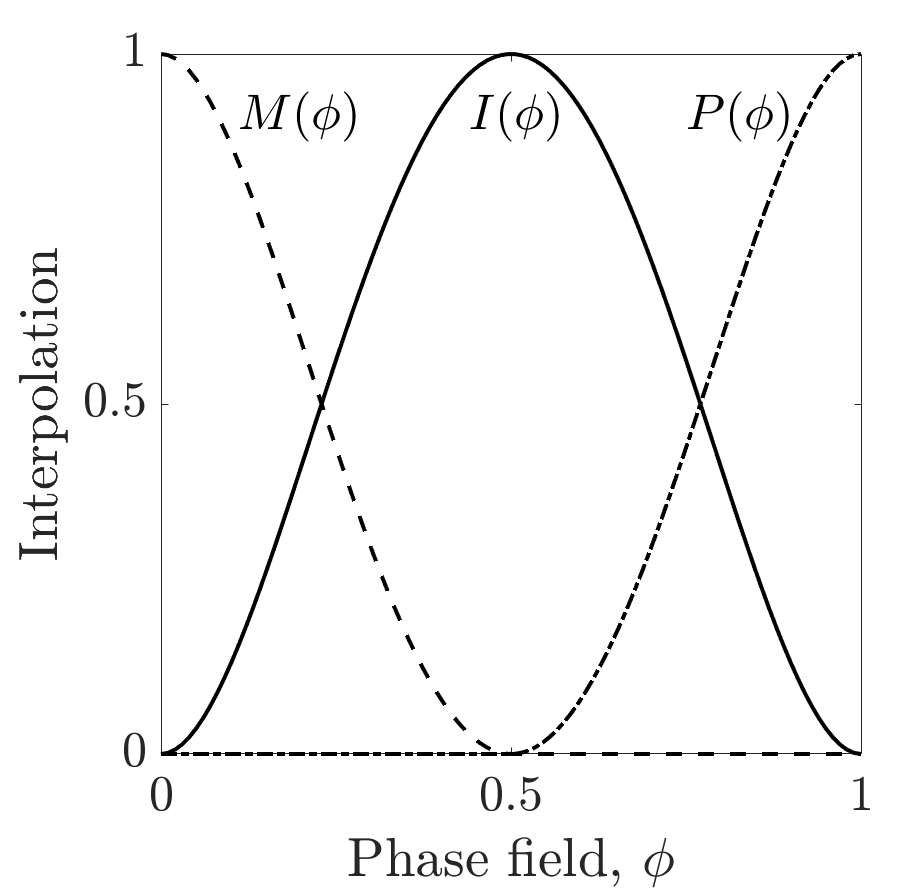}
        \caption*{\hspace{1cm}(a)}
    \end{subfigure} 
    ~ ~    
    \begin{subfigure}[t]{0.45\textwidth}
        \centering
        \includegraphics[width=1\textwidth]{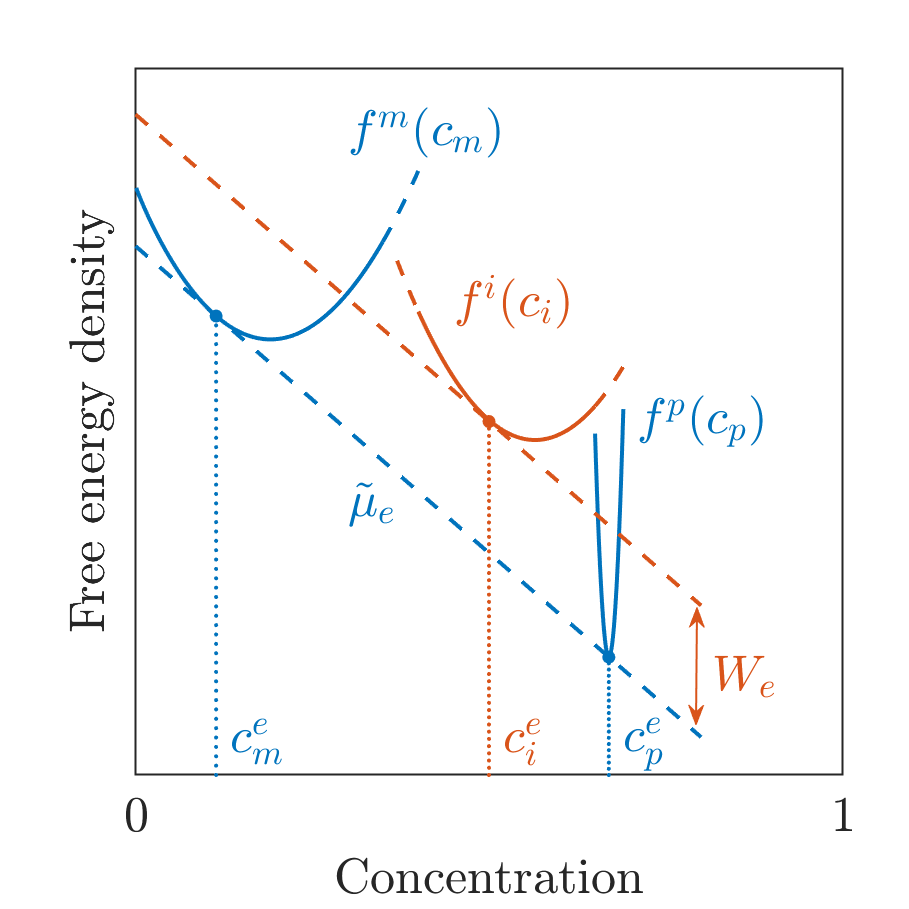}
        \caption*{\hspace{0.5cm}(b)}
    \end{subfigure} 
    \caption{(a) Interpolating functions $M(\phi)$, $I(\phi)$ and $P(\phi)$ corresponding to the matrix ($m$), IB ($i$) and precipitate ($p$) phases, respectively. Phase $i$ is defined at $\phi=0.5$; $m$ at $\phi=0$; $p$ at $\phi=1$. $M$ and $P$ are piecewise, $C^1-$continuous for $\phi\in[0,1]$. $I=16\phi^2(1-\phi)^2$ is a double-well function. (b) Free energy vs. solute concentration curves for $m$, $i$ and $p$. Chemical equilibrium between $m$ and $p$ is given by the common tangent with chemical potential (slope) $\mu_e$; the equilibrium states of $m$ and $p$ are the intersection points of the common tangent with $f^m$ and $f^p$. Equilibrium state of the IB phase is determined by the intersection of the parallel tangent (slope $\mu_e$ and distance $W_e$) to $f^i$.} 
    \label{fig:interpol_fe_schemetic}
\end{figure}

Our modeling framework is focused on material systems that are comprised of matrix ($m$) and precipitate ($p$) phases and IB ($i$) with chemical free energy densities $f^m$, $f^p$, and $f^i$, respectively. Any volume element in the system is described using the concentration fields $c_m(\boldsymbol{x})$, $c_i(\boldsymbol{x})$ and $c_p(\boldsymbol{x})$, representing the matrix, IB, and precipitate, respectively, where $\boldsymbol{x}$ is the position vector. A non-conserved order parameter $\phi(\boldsymbol{x})$ is used to describe the matrix and the precipitate phases. The equilibrium values of $\phi$ are conveniently chosen to be $\phi=0$ and $\phi=1$ at the matrix and precipitate, respectively, and changes smoothly, yet sharply across the IB width. The local free energy density $f(c,\phi)$ is formulated using the interpolating functions $M(\phi)$, $I(\phi)$ and P($\phi$), as
% IB is treated as a phase with a unique function for the free energy density. 
\begin{flalign} \label{eq_kks:f_local}
    f(c,\phi) = M(\phi)f^m(c_m) + I(\phi) f^i(c_i) + P(\phi)f^p(c_p) && \hspace{0.5cm}
\end{flalign} 
where the effective solute concentration $c(\boldsymbol{x})$ is given as a mixture of local phase concentrations as
\begin{flalign} \label{eq_kks:effective_conc_variable}
    c(\boldsymbol{x}) = M(\phi)\,c_m(\boldsymbol{x}) + I(\phi)\,c_i(\boldsymbol{x}) + P(\phi)\,c_p(\boldsymbol{x}) &&
\end{flalign}
At equilibrium, the phase concentrations are constrained by imposing the condition of equal chemical potential (strictly the diffusion potential) between the phases as
\begin{flalign} \label{eq_kks:equal_potential}
    \frac{d f^m (c_m)}{d c_m} = \frac{d f^i (c_i)}{d c_i} = \frac{d f^p(c_p)}{d c_p} \equiv {\mu}(\boldsymbol{x})  &&
\end{flalign}
where $\mu \equiv \mu_B - \mu_A$ is the chemical potential of B measured with respect to that of solvent A in a binary system. We now define the interpolating functions as shown in Fig.\,\ref{fig:interpol_fe_schemetic} such that $M$, $I$ and $P$ describe the exclusive phases $m$ for $\phi=0$, $i$ for $\phi=0.5$ and $p$ for $\phi=1$ respectively through Eqs.\,\ref{eq_kks:f_local} and \ref{eq_kks:effective_conc_variable}. 
% here the height of the barrier potential is set by $f^i$.
This is achieved by setting $M(0)=1$, $I(0.5)=1$, $P(1)=1$ and $M(\phi)+P(\phi)+I(\phi)=1$ for $\phi \in [0,1]$. The center of the IB region is chosen at $\phi=0.5$ for mathematical convenience. We define $I(\phi) = 16\phi^2(1-\phi)^2$.
$M(\phi)$ and $P(\phi)$ $\left( \in C^1[0,1] \right)$ are defined as differentiable, piecewise functions such that matrix side of the IB effectively describes a mixture between $m$ and $i$ given by $M(\phi) = 1 - I(\phi)$ and $P(\phi)=0$; similarly, the precipitate side of the IB $\phi \in (0.5, 1]$ is effectively a mixture between $i$ and $p$ given by $P(\phi) = 1 - I(\phi)$ and $M(\phi)=0$.

The free energy functional for the system at a given temperature $T$ is defined as \cite{wheeler1992phase,kim1999phase}:

\begin{flalign} \label{eq:F_functional}
    F
    = \int_V
    \left[ f(c,\phi;T) + \frac{\varepsilon^2}{2} \left|\nabla \phi
    \right|^2 \right]
\mathrm d V &&
\end{flalign} %\bigintsss_{V}
where the second term in the integrand is the gradient energy density  and $\varepsilon$ is the gradient coefficient associated with the phase field variable $\phi$. 

The time evolution equations for the concentration $c$ and non-conserved  $\phi$ fields follow from the Cahn-Hilliard and Allen-Cahn equations, respectively, as
\begin{flalign} \label{eq:cahn-hilliard}
    \frac{\partial c}{\partial t} = \nabla \cdot \left[M \nabla \left(\frac{\delta F}{\delta c} \right) \right], &&
\end{flalign}
and
\begin{flalign} \label{eq:allen-cahn}
    \frac{\partial \phi}{\partial t} = -L \frac{\delta F}{\delta \phi},  &&
\end{flalign}
where $M$ and $L$ are kinetic parameters related to the atomic and IB mobility, respectively. The equations ensure that the total energy in the system decreases monotonously with time. In the present work, we will deal only with the solutions at steady state equilibrium.

\subsection{Steady-state solutions} \label{subsec:steady-state}

We now consider a one-dimensional system at steady state. The phase field $\phi(x)$ and concentration field $c(x)$ become invariant with time and will be denoted by $\phi_e(x)$ and $c_e(x)$. Steady-state in the evolution equations \ref{eq:cahn-hilliard} and \ref{eq:allen-cahn} results in
\begin{flalign} \label{eq_kks:1D_eq_conc}
    \frac{\delta F}{\delta c} &= \frac{\partial f}{\partial c_e} \equiv {\mu}_e (\text{const.}), \\ \label{eq_kks:1D_eq_phi}
    \frac{\delta F}{\delta \phi} &= \frac{\partial f}{\partial \phi_e} - \varepsilon^2 \frac{d^2 \phi_e}{dx^2} = 0.    &&
\end{flalign}
In Eq.~\ref{eq_kks:1D_eq_conc}, $\mu_e$ is the equilibrium chemical potential, which is a constant across $x$ as required by exchange of atoms between substitutional sites \cite{balluffi2005kinetics}. The constant $\mu_e$ implicitly includes the constraint for conservation of total concentration $c_o$ in the system. 
% (precisely, the diffusion potential or the difference between the chemical potentials of components $\mu_B - \mu_A$ at equilibrium).

Now the condition (Eq.\,\ref{eq_kks:equal_potential}) for equal chemical potential in the phases at $x$ reads
\begin{flalign} \label{eq_kks:equilibrium_potential}
     \frac{df^m(c_m(x))}{dc_m} = \frac{df^i(c_i(x))}{dc_i} = \frac{df^p(c_p(x))}{dc_p} \equiv {\mu}_e (\text{const.})  &&
\end{flalign} %\left.\frac{\partial f}{\partial c_e}\right. =
The above relation implies $c_m(x)=c_m^e$, $c_i(x) = c_i^e$ and $c_p(x) = c_p^e$. That is, the phase concentration fields are constant across space at steady-state. Therefore, the effective concentration field is obtained as
\begin{flalign} \label{eq:conc_profile}
    c_e(x) = M\left(\phi_e(x)\right)\,c_m^e + I\left(\phi_e(x)\right)\,c_i^e + P\left(\phi_e(x)\right)\,c_p^e &&
\end{flalign}
Multiplying Eq.\,\ref{eq_kks:1D_eq_phi} with $d\phi_e/dx$ and integrating piecewise with respect to $x$ from $-\infty$ to $0$, and $0$ to $\infty$ yields (see Appendix\,\ref{appendix:profile}) the equilibrium conditions that determine $c^e_m$, $c^e_i$ and $c^e_p$.
\begin{flalign} \label{eq_kks:pt_eqn_We_def}
    f^i(c_i^e) - f^m(c_m^e) - (c_i^e-c_m^e){\mu}_e = f^i(c_i^e) - f^p(c_p^e) - (c_i^e-c_p^e){\mu}_e \equiv W_e &&
\end{flalign}
which reduces to
\begin{flalign} \label{eq_kks:ct_eqn}
    f^p(c_p^e) - f^m(c_m^e) = (c_p^e - c_m^e){\mu}_e  &&
\end{flalign}

Here, the limits of integration are the far-field matrix ($\phi=0$ as $x\rightarrow\infty$), the IB center ($\phi=0.5$ for $x=0$) and the far-field precipitate ($\phi=1$ as $x\rightarrow \infty$). Eq.\,\ref{eq_kks:ct_eqn} and Eq.\,\ref{eq_kks:equilibrium_potential} constitute the well-known common tangent condition for equilibrium between the bulk phases $m$ and $p$. Given the common tangent, Eq.\,\ref{eq_kks:pt_eqn_We_def} and Eq.\,\ref{eq_kks:equilibrium_potential} represent the parallel tangent condition for equilibrium of the IB phase with the bulk phases. In Eq.\,\ref{eq_kks:pt_eqn_We_def}, the expressions represent the vertical distance between the two tangents, which is defined as $W_e$. For given $f^m$, $f^i$ and $f^p$, the common and parallel tangent constructions determine $c^e_m$, $c^e_i$ and $c^e_p$. This is shown schematically in Fig.\,\ref{fig:interpol_fe_schemetic}b.
The parallel tangent construction is a standard equilibrium condition between a bulk phase and an interface phase in thermodynamic models \cite{hillert1975lectures}, which has been widely used in the context of free surfaces and GBs \cite{wynblatt1979interfacial,howe1997interfaces, abdeljawad2017grain}. % GB phase field segregation parallel tangent

Using Eqs.\,\ref{eq_kks:f_local} and \ref{eq_kks:pt_eqn_We_def}, the equilibrium phase-field Eq.\,\ref{eq_kks:1D_eq_phi} can be conveniently reduced (Appendix\,\ref{appendix:profile}) to the form 
\begin{flalign} 
    \varepsilon^2\frac{d^2\phi_e}{dx^2} = W_e\frac{dI(\phi_e)}{d\phi_e}  &&
\end{flalign}
Multiplying by $d\phi_e/dx$ and integrating gives
\begin{flalign} \label{eq_kks:eq_phase-field_final}
   \frac{\varepsilon^2}{2} \left(\frac{d\phi_e}{dx}\right)^2 = W_eI(\phi_e) &&
\end{flalign}
The above relation is a typical steady-state result in diffuse-interface models. It represents the phase-field profile that minimizes the excess (IB) energy in the system arising through contributions from the gradient energy (left side of the equation) and the barrier potential (right side). Eq.\,\ref{eq_kks:eq_phase-field_final} depends only on the functional form of $I$, which is chosen as the standard double-well form $16\phi^2(1-\phi)^2$. While $W_e$ is concentration-dependent (via $c^e_m$,$c^e_i$ and $c^e_p$) it is constant across space, and therefore the equation can be analytically integrated to yield the steady-state solution for the phase-field as
\begin{flalign} \label{eq:hyperbolic_tangent}
    \phi_e(x) = \frac{1}{2}\left[1+\tanh\left(\frac{2\sqrt{2W_e}}{\varepsilon}x\right)\right]   &&
\end{flalign}
The diffuse IB width $\lambda$ can be measured as the spatial distance defined by $\phi_e$, say between $\phi_e = 0.1$ and $\phi_e=0.9$ (bounds chosen for convenience of analytic integration). This gives
\begin{flalign} \label{eq_kks:width}
    \lambda = \int_{0.1}^{0.9} \frac{dx}{d\phi_e} d\phi_e \approx \frac{1.1\varepsilon}{\sqrt{2W_e}}  &&
\end{flalign}

\subsection{Excess properties and Gibbs adsorption} \label{subsec:excess_GA}
In Gibbs thermodynamics, the IB is an infinitesimally thin layer that separates homogeneous bulk phases. This mathematical interface is associated with the excess IB energy $\gamma$ and excess solute concentration contained in the real, diffuse-interface system \cite{sutton1995interfaces}. To obtain these excess quantities from the steady-state phase-field solution, the Gibbs dividing surface is introduced at $x=0$. $\gamma$ is then evaluated as the excess energy per unit IB area $A_i$, and is derived in  Appendix\,\ref{appendix:IB_energy} as
\begin{flalign} \label{eq:gamma_int}
    \gamma = \int_{-l}^{l} 2W_eI(\phi_e(x)) dx \approx \frac{2\varepsilon \sqrt{2W_e}}{3} &&
\end{flalign}
where $l$ is chosen far from the interfacial region; the integrand $\Omega(x) \equiv  2W_eI(\phi_e(x))$ is the excess grand potential \cite{provatas2011phase}. $\gamma$ is a function of the IB concentration through $W_e$. The integrand vanishes in the bulk phases as $l\rightarrow \pm \infty$ due to the common tangent construction. Therefore, $\gamma$ converges and is independent of the dividing surface convention. Using Eq.\,\ref{eq_kks:width}, we can eliminate $\varepsilon$ and express the IB energy in terms of the concentration-dependent parallel tangent distance and IB width as $\gamma \approx 1.2 W_e \lambda$.

The excess solute concentration $C_{xs}$, evaluated with respect to the dividing surface at $x=0$, is obtained (Appendix\,\ref{appendix:IB_energy}) as $C_{xs} \approx 0.3(2c_i^e - c_m^e - c_p^e) \lambda$.
For homophase boundaries $C_{xs}$ becomes independent of the Gibbs dividing surface location since $c^e_p=c^e_m$. However for an IB $C_{xs}$ is a function of the dividing surface location since $c^e_p \neq c^e_m$ \cite{johnson1979interfacial}. 
Therefore, a measure of solute excess independent of the choice of the dividing surface convention is the appropriate thermodynamic quantity for IBs \cite{cahn1979interfacial}. The method to obtain Gibbs adsorption equation with invariant thermodynamic quantities was developed by Cahn \cite{cahn1979interfacial} and adapted to diffuse interface models by McFadden and Wheeler \cite{mcfadden2002gibbs}. Following \cite{mcfadden2002gibbs}, the invariant form of solute excess $\Gamma_{xs}$ is given by (Appendix \ref{appendix:GA})
\begin{flalign} \label{eq:inv_solute_xs}
    \Gamma_{xs} = \int_{-l}^{l} \left[(c_e(x)-c^e_m)  - \frac{(c^e_p-c^e_m)}{(s^p(c^e_p)-s^m(c^e_m))}(s_e(x)-s^m(c^e_m)) \right] dx &&
\end{flalign}
here, $s^r = df^r(c^e_r)/dT$ is the configurational entropy density of the bulk phase ($r=m,p$) at equilibrium concentration $c^e_r$; $s_e(x)$ and $c_e(x)$ are the entropy and solute concentration profiles of the diffuse interface system at equilibrium. Within the $m$ and $p$ regions far from the IB, the integrand vanishes. $\Gamma_{xs}$ therefore converges as $l\rightarrow \pm \infty$, making it an invariant quantity. $\Gamma_{xs}$ represents the excess concentration in the diffuse system over what would be present in a comparison system of homogeneous $m$ and $p$ phases containing the same volume and entropy \cite{cahn1979interfacial,umantsev2001continuum}.  Eq.\,\ref{eq:inv_solute_xs} can also be integrated analytically and is given by Eq.\,\ref{eq:anal_inv_excess}. For the binary system considered in our work, the IB energy $\gamma$ can be regarded as a function of temperature $T$ alone, and the relevant Gibbs adsorption equation \cite{mcfadden2002gibbs} is given by 
\begin{flalign} \label{eq_kks:gibbs_ads}
    \frac{d\gamma}{dT} = - \Gamma_{xs} \frac{d{\mu}_e}{dT} 
     && %= -\int_{-l}^{l} \left[(s(x)-s^m(c^e_m)) - \frac{(c(x)-c^e_m)}{(c^e_p-c^e_m)}(s^p(c^e_p)-s^m(c^e_m)) \right] dx
\end{flalign} 
where $d\mu_e/dT$ captures the change in the bulk phase equilibrium with $T$ in the two phase coexistence region of the phase diagram. The applicability of Eqs.\,\ref{eq:inv_solute_xs} and \ref{eq_kks:gibbs_ads} in the context of our model is shown via analytic derivation in Appendix \ref{appendix:GA}, and through parametric study in Sec.\,\ref{param_study:GA}.
\FloatBarrier
\section{Solution thermodynamics} \label{sec:regular_soln}

In this section we study the IB properties arising from our phase-field formulation. We model the energetics of the IB phase and the bulk phases by assuming the regular solution behavior. 
The regular solution free energy characteristic to a phase $r$ (where $r = m, p, i$) with concentration $c_r$ is given by
\begin{flalign} \label{eq:RS_phase_FE}
    f^r(c_r) &= G_A^r (1-c_r) + G_B^r c_r + L^r (1-c_r) c_r + \frac{RT}{v_m} \left((1-c_r) \ln{(1-c_r) + c_r \ln{c_r}}\right) &&
\end{flalign}
where $G_A^r$ and $G_B^r$ are the free energies of the pure components in phase $r$; $L^r$ is the regular solution interaction parameter which describes the non-ideal interaction between $A$ and $B$ in phase $r$; the last term is the ideal configurational entropy for mixing of $A$ and $B$ atoms; $v_m$ is the molar volume, which is assumed as a constant.

Using the regular solution free energy for the IB phase in $d f^i(c^e_i)/d c^e_i = \mu_e$  (Eq.\,\ref{eq_kks:equilibrium_potential}), the equilibrium concentration of the IB phase $c^e_i$ is obtained as  
\begin{flalign} \label{eq:cie_parallel_tgt}
    c^e_i = \frac{1}{1+\exp{\left(\frac{G^i_B-G^i_A+L^i-2L^i c^e_i - \mu_e}{RT/v_m}\right)}}  &&
\end{flalign} % Note: changed sign within exp
% = \frac{(1-c^e_i)c^e_m}{(1-c^e_m)} \exp{\left(\frac{G^i_A-G^m_A-G^i_B+G^m_B-L^i+L^m+2L^ic^e_i-2L^mc^e_m}{RT/v_m}\right)}
where $\mu_e$ is known from the common tangent between the bulk phase free energies. For a given bulk system and $T$, $\mu_e$ is fixed, and Eq.\,\ref{eq:cie_parallel_tgt} can be used to determine $c^e_i$ for various IB phases defined by $(G^i_A,G^i_B,L^i)$. For an ideal solution of non-interacting atoms ($L^i = 0$), Eq.\,\ref{eq:cie_parallel_tgt} reduces to a Fermi-Dirac distribution of $c^e_i$ over the energy states $G^i_B-G^i_A$. Therefore $G^i_B-G^i_A << {\mu}_e$ yields $c^e_i \rightarrow 1$, $G^i_B-G^i_A >> {\mu}_e$ yields $c^e_i \rightarrow 0$, and $c^e_i = 0.5$ yields $G^i_B - G^i_A = {\mu}_e$. These will be used as a guide for the parametric study in the next section to set IB parameters that result in a strong IB phase concentration, $c^e_i>\max(c^e_m,c^e_p)$.
\FloatBarrier

% \section{Regular solution approximation}

\subsection{Parameterization}

We non-dimensionalize the steady-state equations by setting $f^r=\left(RT_o/v_m\right)\tilde{f}^r$ and $x=l_o\tilde{x}$. Here, $T_o$ is the characteristic temperature, $RT_o/v_m$ is the characteristic energy, $l_o$ is the characteristic length scale of the system, and $r = m,i,p$. The non-dimensionalized free-energy can be written as: $\tilde{f}^r(c_r) = \tilde{G}^r_A(1-c_r) + \tilde{G}^r_B c_r + \tilde{L}^r(1-c_r)c_r + \tilde{T} \left((1-c_r)\ln{(1-c_r)} + c_r\ln{c_r} \right)$, where $\tilde{T}=T/T_o$. The non-dimensional quantities are denoted by the tilde symbol over the corresponding dimensional quantities. The resulting steady-state equations have the same form as Eqs.~\ref{eq_kks:1D_eq_conc} and \ref{eq_kks:1D_eq_phi} but replaced with non-dimensional quantities $\tilde{\varepsilon}^2=\varepsilon^2/\left[l_o^2\left(RT_o/v_m\right)\right]$ and $\tilde{\mu}_e = \mu_e/\left(RT_o/v_m\right)$. 
% $\tilde{\varepsilon}^2 = 3$ throughout this study.

\begin{figure}[h!]
    \centering
    \includegraphics[width=0.5\textwidth]{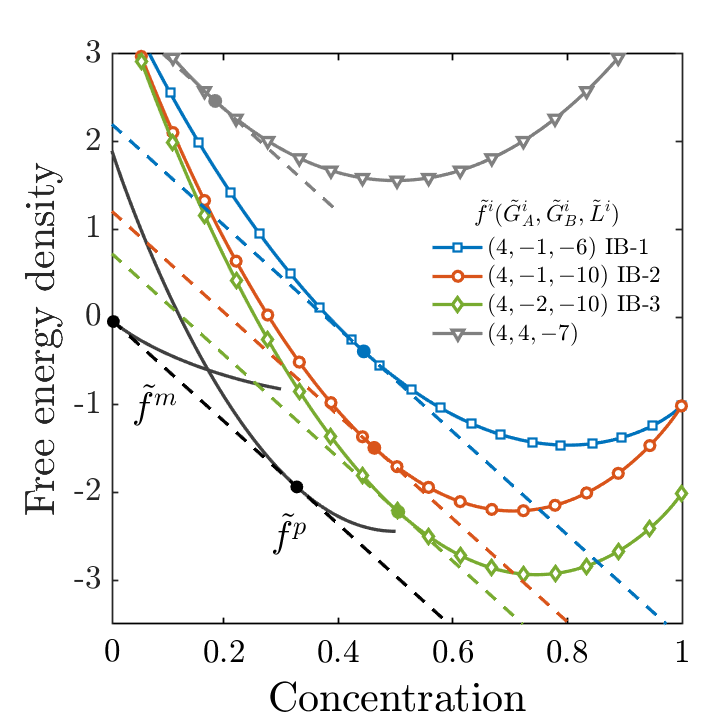}
    \caption{Free energy density ($\tilde{f}$) versus concentration for matrix ($m$), precipitate ($p$) and three different IB ($i$) phases. Regular solution parameters for the bulk phases are: $(\tilde{G}^m_A,\tilde{G}^m_B,\tilde{L}^m)=(0,0,-1)$, $(\tilde{G}^p_A,\tilde{G}^p_B,\tilde{L}^p)=(2,2,-15)$. The equilibrium state of the bulk phases is given by the common tangent construction. The state of an IB phase in equilibrium with the bulk is obtained via the parallel tangent construction and the equilibrium states are marked by the dots on the curves.}
    \label{fig:fe_parametric}
\end{figure}

To illustrate the steady-state solutions, we choose the following regular solution parameters: $(\tilde{G}^m_A,\tilde{G}^m_B,\tilde{L}^m) = (0,0,-1)$, $(\tilde{G}^p_A,\tilde{G}^p_B,\tilde{L}^p) = (2,2,-15)$. 
These parameters are chosen so that, at $\tilde{T}=1$, we obtain a two-phase equilibrium with negative $\tilde{\mu}_e$ ($\approx -6$) (slope of common tangent in Fig.~\ref{fig:fe_parametric}). The matrix exhibits lean solute solubility, i.e. $c^e_m \approx 0.07$, and the precipitate has intermediate solute solubility of  $c^e_p \approx 0.32$. The phase diagram and the coexistence region $m+p$ for the chosen bulk system is shown in inset of Fig.~\ref{fig:fe_vs_T_Gibbs_ads}. The bulk system modeled here is representative of a realistic two-phase region consisting of a terminal solid-solution and an intermediate intermetallic phase \cite{gaskell2017introduction,hillert2007phase}. Partitioning of $B$ to the precipitate is captured by the strong $A$--$B$ interaction in the precipitate $\tilde{L}^p \ll \tilde{L}^m$. 

Free energy density plots are shown in Fig.~\ref{fig:fe_parametric} for three IB phases $(\tilde{G}^i_A,\tilde{G}^i_B,\tilde{L}^i)$ that will be used to evaluate the steady-state profiles in the next section. The parameters are: $(4,-1,-6)$ for IB-1, $(4,-1,-10)$ for IB-2 and $(4,-2,-10)$ for IB-3. The equilibrium states of the IBs are obtained from the parallel tangent construction. The difference between the equilibrium IB concentrations ($c^i_e$) of the different IBs can be understood from Eq.~\ref{eq:cie_parallel_tgt} in terms of the ideal solution parameters $\tilde{G}^i_B-\tilde{G}^i_A$, the effect of non-ideal interaction $\tilde{L}^i$, the bulk phase equilibrium $\tilde{\mu}_e$ and the extent of entropic influence governed by $\tilde{T}$. The magnitudes of $\tilde{L}^i$ used here signify an $A$--$B$ interaction strength intermediate to that of $m$ and $p$. The large positive value for $\tilde{G}^i_A$ represents the unfavorable energetics for pure component $A$ at the IB. The chosen parameters satisfy $\tilde{G}^i_B-\tilde{G}^i_A \approx \tilde{\mu}_e$ and $\tilde{L}^i < 0$, which ensures favorable $A$--$B$ energetics with $c^e_i \approx 0.5$.
% Concentration of $B$ at the IB is predominantly driven by: lower pure component energy for $B$ relative to $A$ ($G^i_B<\tilde{G}^i_A)$.

\subsection{Steady state profiles} \label{sec:param_study:steady-state}

\begin{figure}[h!]
    \centering
    \begin{subfigure}[t]{0.5\textwidth}
        \centering
        \includegraphics[width=1\textwidth]{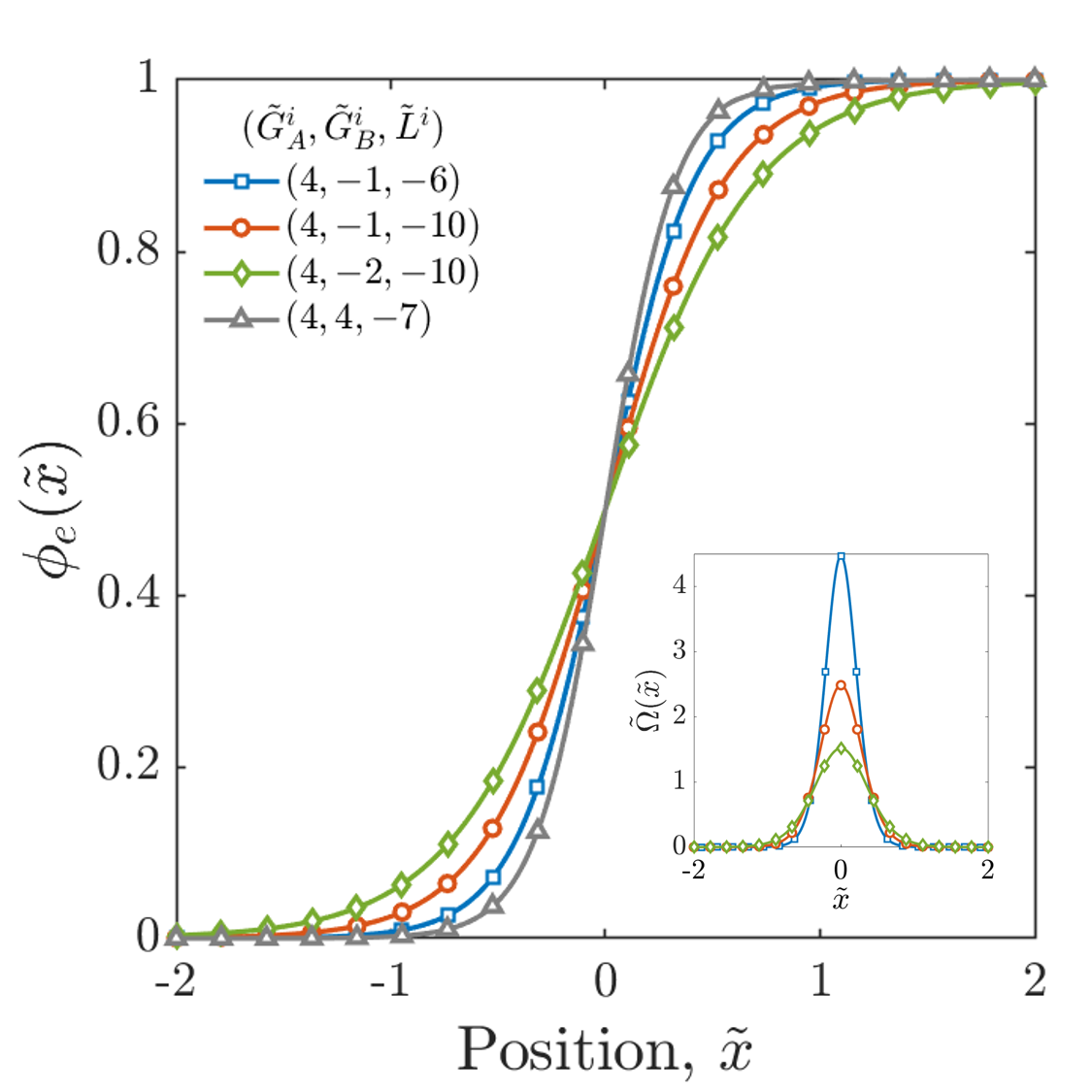}
        % \caption{Lorem ipsum}
    \end{subfigure}%
    ~ ~    
    \begin{subfigure}[t]{0.5\textwidth}
        \centering
        \includegraphics[width=1\textwidth]{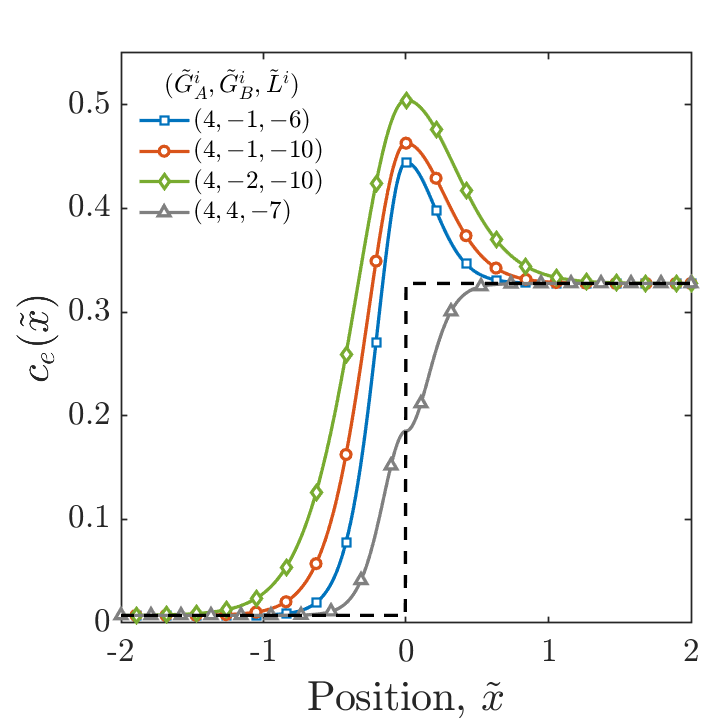}
        % \caption{Lorem ipsum}
    \end{subfigure}%
    \caption{1D steady-state profiles for (a) phase-field $\phi_e(\tilde{x})$ and (b) concentration for different IBs in the bulk system given in Fig.~\ref{fig:fe_parametric}). $\tilde{\varepsilon}^2=3$. Note the larger diffuse-interface widths for IB with smaller parallel tangent distance (or barrier height $\tilde{W}_e$). Interface center concentration ($\tilde{x}=0$) is given by the equilibrium IB phase concentration in Fig.~\ref{fig:fe_parametric}. Excess interface concentration can be seen to be a function of concentration at $\tilde{x}=0$ and diffuse IB width. Inset (a): the local excess potential $\tilde{\Omega}$ (Eq.~\ref{eq:gamma_int}) across the diffuse IB.} 
    \label{fig:phie_ce_profiles}
\end{figure}

The steady-state diffuse-interface profiles for different IB phases are shown in Fig.~\ref{fig:phie_ce_profiles}. The diffuseness or width $\tilde{\lambda}$ of the phase-field profile $\phi_e(\tilde{x})$ (Fig.~\ref{fig:phie_ce_profiles}a) increases from IB-1 to IB-3. The diffuse-interface properties are determined by the parallel tangent distance $\tilde{W}_e$ (Eq.~\ref{eq:cie_parallel_tgt}), which decreases from IB-1 to IB-3 (as shown in Fig.~\ref{fig:fe_parametric}). Since $\tilde{\varepsilon}^2 = 3$ for all IBs, the variation in width results from its inverse dependence with $\sqrt{\tilde{W}_e}$ (Eq.~\ref{eq_kks:width}). The concentration profiles $c_e(x)$, (Fig.~\ref{fig:phie_ce_profiles}b) show a non-monotonous variation in the solute concentration across the system given by Eq.~\ref{eq:conc_profile}. The concentrations far-field of the IB correspond to $c^e_m$ as $\tilde{x}\rightarrow-\infty$ and $c^e_p$ as $\tilde{x}\rightarrow+\infty$, in accordance with that given by the common tangent. The peak or interface-center ($\tilde{x}=0$) concentrations for the different IBs correspond to the exclusive IB phase concentrations $c^i_e$ as given by the parallel tangents in Fig.~\ref{fig:fe_parametric}. The width of the concentration profiles are related to the width of the phase-field profiles as shown in Eq.~\ref{eq:conc_profile}. Therefore, in addition to the IB phase concentration $c^i_e$, the total solute content in the diffuse IB region is governed by the barrier height $\tilde{W}_e$ and the gradient energy coefficient $\tilde{\varepsilon}^2$. 

For an IB phase given by $(\tilde{G}^i_A,\tilde{G}^i_B,\tilde{L}^i)=(4,4,-7)$ (Fig.~\ref{fig:fe_parametric}), the concentration profile is characterized by a step at $\tilde{x}=0$. For these parameters, there is no pure-component energetic preference for either $A$ or $B$ in any of the phases since $G^r_B-G^r_A=0$ ($r=m,i,p$), and therefore the ordering of the $A$--$B$ interaction strength $\tilde{L}^p<\tilde{L}^i<\tilde{L}^m$ results in $c^e_p>c^e_i>c^e_m$. Traditional phase-field models generally yield a sigmoidal concentration profile between the matrix and precipitate phases. The present description is characterized by a unique peak concentration (for $c^e_i>c^e_p$) or an intermediate step (for $c^e_m<c^e_i<c^e_p$) resulting from the inclusion of an IB phase.

The excess grand potential $\tilde{\Omega}(\tilde{x}) = 2W_eI(\phi_e(\tilde{x}))$, across the diffuse interface is shown in the inset of Fig.~\ref{fig:phie_ce_profiles}a. The height of the potential is set by $\tilde{W}_e$ and therefore follows IB-1$>$IB-2$>$IB-3. The integral under these curves yield the excess (IB) energy $\gamma$ in the system via Eq.~\ref{eq:gamma_int}. The IB energy can be dimensionalized as $\gamma = \left(RT_ol_o/v_m\right)\tilde{\gamma}$. Assuming $v_m = 7 \times 10^{-6}$ m$^3$/mol, and setting $T_o = 300$ K  and $l_o = 1$ nm yields $\gamma \approx 0.36~\tilde{\gamma} $ J/m$^2$. Therefore, $\gamma$ lies in the range $0.2-1$ J/m$^2$, which is the range typically associated with semicoherent IBs \cite{howe1997interfaces}.

\subsection{Excess properties}

\begin{figure}[h!]
    \centering
    \begin{subfigure}[t]{0.5\textwidth}
        \centering
        \includegraphics[width=1\textwidth]{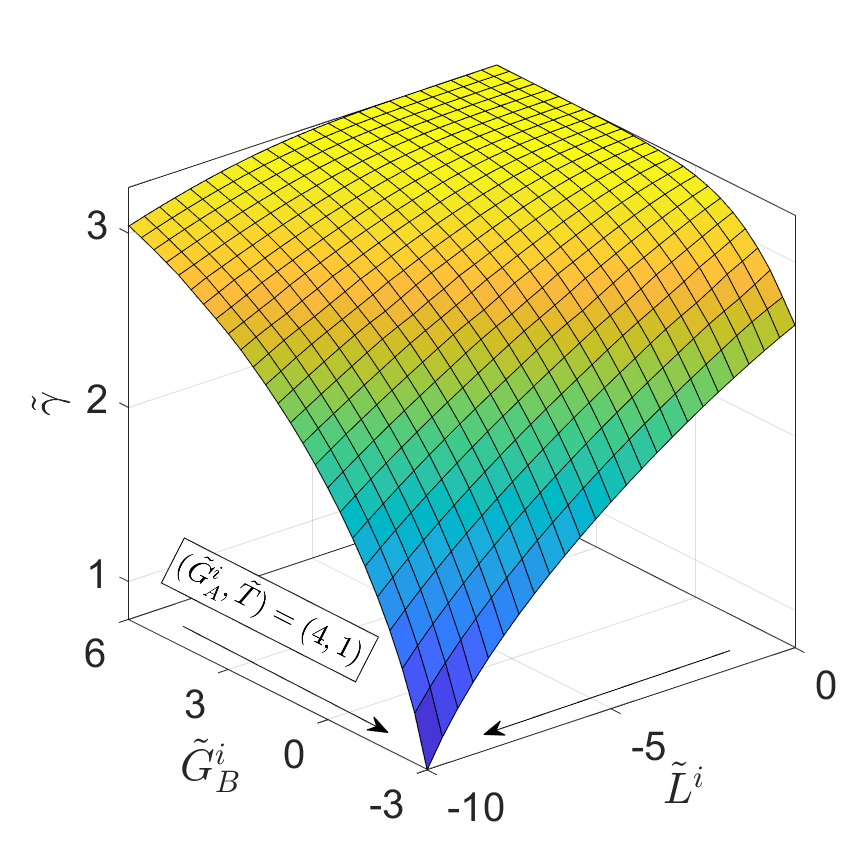}
        % \caption{Lorem ipsum}
    \end{subfigure}%
    ~ ~    
    \begin{subfigure}[t]{0.5\textwidth}
        \centering
        \includegraphics[width=1\textwidth]{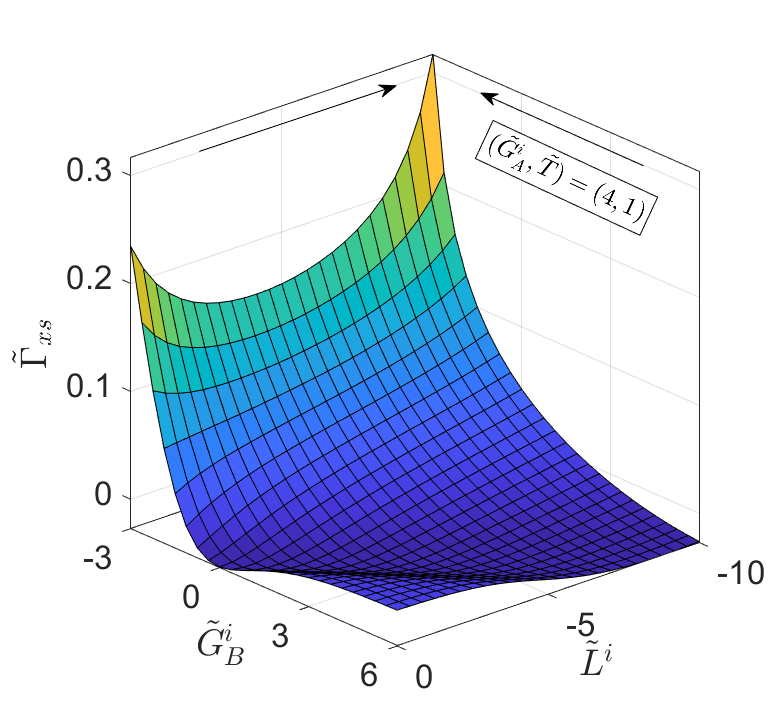}
        % \caption{Lorem ipsum}
    \end{subfigure}%
    \caption{Surface plots of excess IB properties as a function of IB parameters $(\tilde{G}^i_B,\tilde{L}^i)$. (a) Dependence of IB energy $\tilde{\gamma}$ (from Eq.~\ref{eq:gamma_int}) and (b) excess solute concentration $\tilde{\Gamma}_{xs}$ (from Eq.~\ref{eq:inv_solute_xs}) for IBs $(\tilde{G}^i_B,\tilde{L}^i)$ in equilibrium with the bulk phases shown in Fig.~\ref{fig:fe_parametric}. $(\tilde{G}^i_A, \tilde{T}, \tilde{\varepsilon}^2) = (4, 1, 3)$. IB phases with lower pure component energy for $B$ ($\tilde{G}^i_B$) and more favorable $A$--$B$ interaction ($\tilde{L}^i$) exhibit lower $\tilde{\gamma}$. (The $\sim$ symbol denotes non-dimensional quantities.)} 
    \label{fig:excess_RS_params}
\end{figure}

The IB energy $\tilde{\gamma}$ and the invariant solute excess $\tilde{\Gamma}_{xs}$ are obtained from the diffuse interface system with reference to the flat interface description of the classical thermodynamic system (Sec.~\ref{subsec:excess_GA}). Surface plots demonstrating the dependence of $\tilde{\gamma}$ and $\tilde{\Gamma}_{xs}$ as a function of the IB phase parameters $(\tilde{G}^i_B,\tilde{L}^i)$ are presented in Fig.~\ref{fig:excess_RS_params} for the bulk system in Fig.~\ref{fig:fe_parametric} and constant IB parameters $(\tilde{G}^i_A,\tilde{\varepsilon}^2) = (4,3)$.
% (Appendix) giving the IB energy $\gamma$ (Eq.~\ref{eq:gamma_int}) and the invariant solute excess $\Gamma_{xs}$ (Eq.~\ref{eq:inv_solute_xs}). 
For large $\tilde{G}^i_B$ (i.e. unfavorable energetics for pure $B$ relative to $A$ at $i$) and weak $\tilde{L}^i$ (ideal mixing), $\tilde{\gamma}$ plateaus to a maximum value. This corresponds to an IB concentration $c^e_i \rightarrow 0$. Decreasing $\tilde{G}^i_B$ (i.e. making the presence of $B$ at the IB energetically favorable) or $\tilde{L}^i$ (i.e. strengthening $A$--$B$ interaction at IB) results in a decrease in $\tilde{\gamma}$. For $\tilde{G}^i_B \ll \tilde{G}^i_A$ and $\tilde{L}^i \ll 0$, $\tilde{\gamma}$ reduces steeply towards the zero value; this corresponds to $c^e_i \rightarrow 1$ and a large $\tilde{\Gamma}_{xs}$. 
$\tilde{\Gamma}_{xs}$ is a function of the concentration and entropy profiles (Eq.~\ref{eq:inv_solute_xs}) and the surface plot reflects the variation $-d\tilde{\gamma}/d\tilde{\mu}_e(\tilde{T})$ over the IB parameter space.

% By decreasing $\tilde{G}^i_B$ (relative to $\tilde{G}^i_A$), the presence of $B$ at the IB phase is made energetically more favorable. Therefore, an increase in IB phase concentration $c^e_i$, and a corresponding decrease in the height of the barrier potential $\tilde{W}_e$ is expected. The excess properties reflect these trends, as observed by the increase in the excess solute $\tilde{\Gamma}_{xs}$ and a decrease in the IB energy $\tilde{\gamma}$ ($\propto \sqrt{\tilde{W}_e}$) in Fig.~\ref{fig:excess_RS_params}. However, the contribution to $\tilde{\Gamma}_{xs}$ arises from the IB phase concentration $c^e_i$ directly and from the IB width $\tilde{\lambda}$ indirectly. This can be readily seen from $\tilde{C}_{xs} \propto (2c^e_i-c^e_m-c^e_p)\tilde{\lambda}$ (or from the complete form in \ref{eq:anal_inv_excess}), where $\tilde{W}_e$ governs the IB width via $\tilde{\lambda} \propto 1/\sqrt{\tilde{W}_e}$.

%  Like $\tilde{G}^i_B$, $\tilde{L}^i$ effects a similar but minor variation on $\tilde{\gamma}$ and solute excess.  Since $c^e_i$ is already close to equiatomic for given $\tilde{G}^i_B-\tilde{G}^i_A$ and $\tilde{\mu}_e$,  further strengthening $A$--$B$ interface interaction doesn't alter $c^e_i$ significantly but produces lower IB energetics captured by $\tilde{W}_e$.
%  Decreasing IB energy of $B$ ($\tilde{G}^i_B$) and increasing the strength of $A$--$B$ interface interaction ($\tilde{L}^i$) together results in a steep increase in solute excess (and width) and a corresponding decrease in $\tilde{\gamma}$.

\subsection{Temperature dependence and Gibbs adsorption} \label{param_study:GA}

In this section, we present the effect of temperature $\tilde{T}$ on the equilibrium state ($c^e_i$) of the IB phase and the excess properties ($\tilde{\gamma}$,$\tilde{\Gamma}_{xs}$) arising from the IB phase and its gradients with the bulk phases. Variation in $\tilde{T}$ will cause a shift in the bulk-phase equilibrium given by $\tilde{\mu}_e(\tilde{T})$, which will effect a variation on the IB state ($c^e_i$, $W_e$), the  diffuse-interface profiles $c_e(\tilde{x})$, $\phi_e(\tilde{x})$, and on the excess quantities $\tilde{\gamma}$ and $\tilde{\Gamma}_{xs}$.

\begin{figure}[h!]
    \centering
    \begin{subfigure}[t]{0.5\textwidth}
        \centering
        \includegraphics[width=0.95\textwidth]{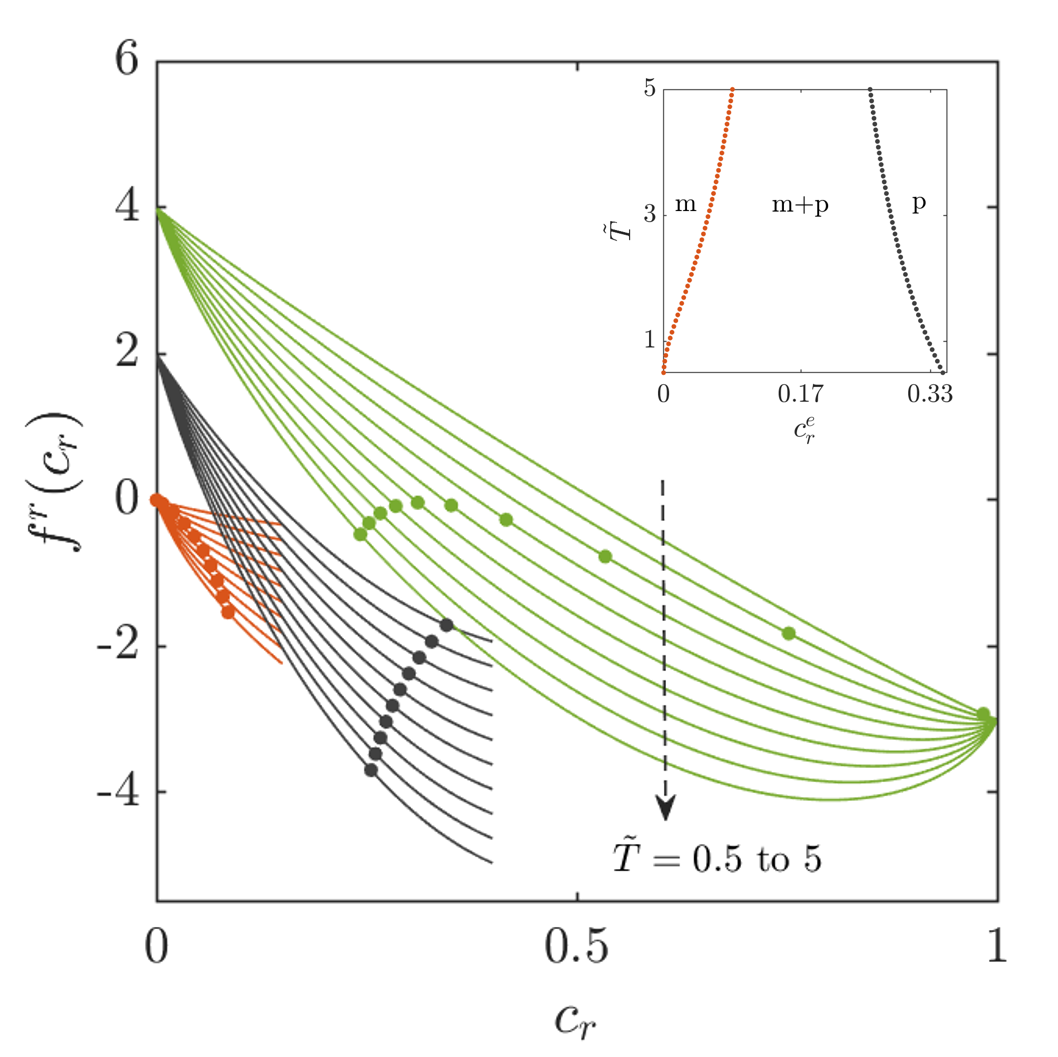}
        % \caption{Lorem ipsum}
    \end{subfigure}%
    ~ ~    
    \begin{subfigure}[t]{0.5\textwidth}
        \centering
        \includegraphics[width=1\textwidth]{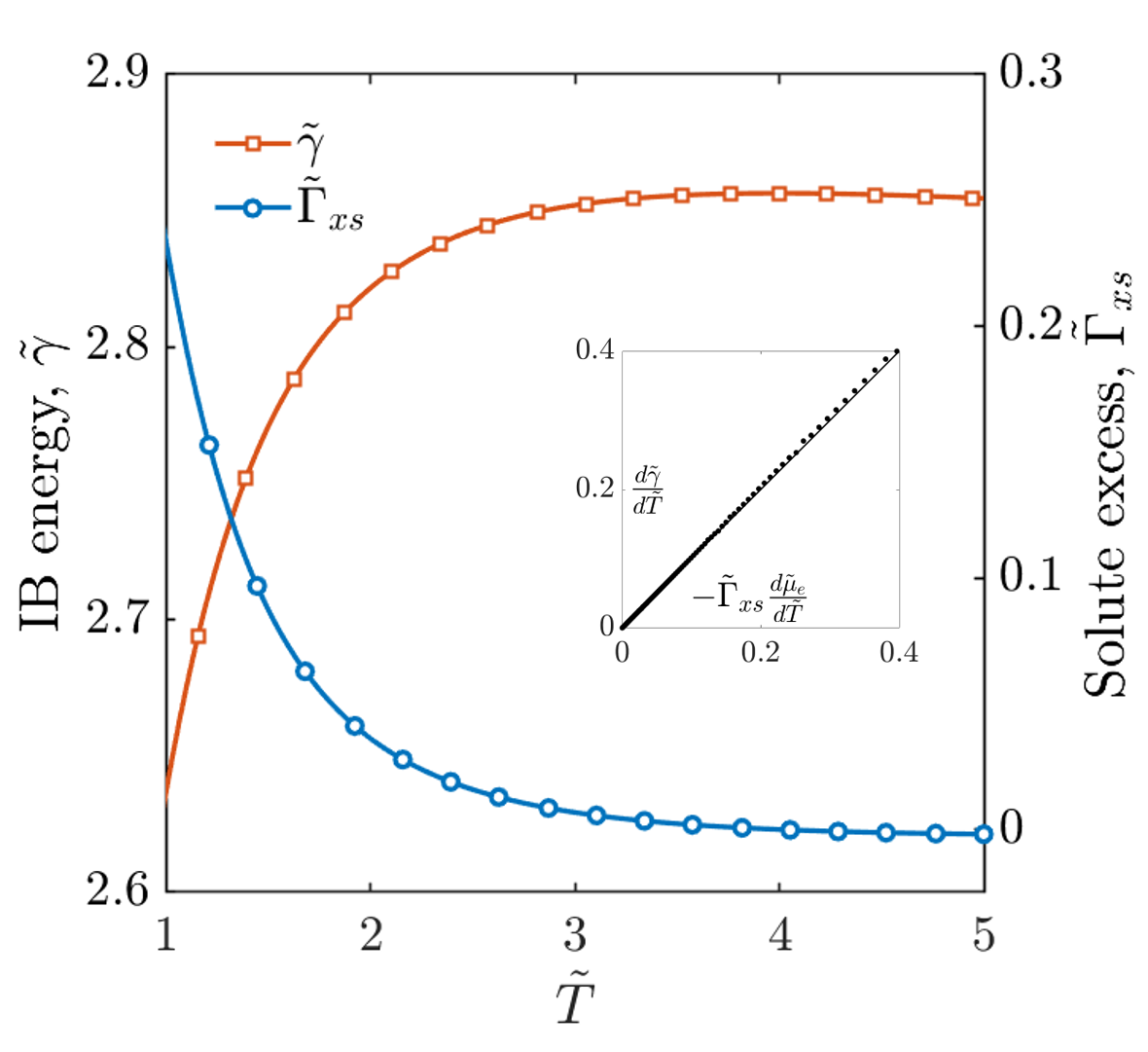}
        % \caption{Lorem ipsum}
    \end{subfigure}%
    \caption{(a) Free energy density versus concentration curves for a hypothetical system composed of matrix $(\tilde{G}^m_A,\tilde{G}^m_B,\tilde{L}^m)=(0,0,-1)$, precipitate $(\tilde{G}^p_A,\tilde{G}^p_B,\tilde{L}^p)=(2,2,-15)$ and IB phase $(\tilde{G}^i_A,\tilde{G}^i_B,\tilde{L}^i)=(4,-3,0)$ for temperatures $\tilde{T}=0.5$ to $5$. Filled marker on an IB free energy curve represents the IB state in equilibrium with the bulk states (markers on $f^m$ and $f^p$ curves) at the corresponding $\tilde{T}$. Inset: the bulk phase diagram. (b) Variation in IB energy $\tilde{\gamma}$ (from Eq.~\ref{eq:gamma_int}) and solute excess $\tilde{\Gamma}_{xs}$ (from Eq.~\ref{eq:inv_solute_xs}) with $\tilde{T}$ obtained from the diffuse-interface model ($\tilde{\varepsilon}=5$) for the hypothetical $m$-$i$-$p$ system described in (a). Inset: Gibbs adsorption equation $\frac{d\tilde{\gamma}}{d\tilde{T}}=-\tilde{\Gamma}_{xs}\frac{d\tilde{\mu}_e}{d\tilde{T}}$ (Eq.~\ref{eq_kks:gibbs_ads}) \cite{mcfadden2002gibbs} is shown to hold.} 
    \label{fig:fe_vs_T_Gibbs_ads}
\end{figure}

The variation in the free energy curves $f^m$, $f^i$ and $f^p$ with  $\tilde{T}$ is shown in Fig.~\ref{fig:fe_vs_T_Gibbs_ads}. Corresponding equilibrium states of the bulk phases ($c^e_m$,$c^e_p$) and the IB phase ($c^e_i$) obtained from the common and parallel tangent constructions, respectively, are marked on the free energy curves. The phase diagram and the coexistence region $m+p$ for the chosen bulk system is shown in inset of Fig.~\ref{fig:fe_vs_T_Gibbs_ads}. 

\FloatBarrier

We choose a hypothetical case of an ideal IB phase $(\tilde{G}^i_A,\tilde{G}^i_B,\tilde{L}^i) = (4,-3,0)$ and demonstrate its Gibbs adsorption behavior. The parameters favor $c^e_i \rightarrow 1$ at low $\tilde{T}$. With an increase in $\tilde{T}$ the effect of entropy becomes more prominent, causing the mixing of $A$ and $B$ to become more favorable in each of the phases (Fig.~\ref{fig:fe_vs_T_Gibbs_ads}a).
% The free energy curves in Fig.~\ref{fig:fe_vs_T_Gibbs_ads}a show that 
Since all phases of the system, including the IB phase, are open to exchanging solute between each other, the solute desegregates from the IB and redistributes towards a more uniform distribution throughout the system. At low temperature ($\tilde{T}=1$), the energetic preference for component $B$ at the IB results in $c^e_i\rightarrow 1$, whereas at higher temperatures ($\tilde{T}=5$), the entropic contribution results in $c^e_m<c^e_i<c^e_p$. Correspondingly, $\tilde{\Gamma}_{xs}$ (Fig.~\ref{fig:fe_vs_T_Gibbs_ads}b) decreases with $\tilde{T}$ and $\tilde{\Gamma}_{xs} \rightarrow 0$ at high $\tilde{T}$. On the other hand, $\gamma$ increases but with a decreasing slope $d\tilde{\gamma}/d\tilde{T}$ and approaches a constant value at high $\tilde{T}$. The observed relationship between $\tilde{\Gamma}_{xs}$ and $d\tilde{\gamma}/d\tilde{T}$ is in agreement with the Gibbs adsorption relation (Eq.~\ref{eq_kks:gibbs_ads}). 

The applicability of the invariant solute excess $\tilde{\Gamma}_{xs}$, and the compatibility of our diffuse interface formulation with the Gibbs adsorption relation is demonstrated by Fig.~\ref{fig:fe_vs_T_Gibbs_ads}b(inset). 
The excess properties $\tilde{\Gamma}_{xs}$ (adapted from \cite{mcfadden2002gibbs}) and $\tilde{\gamma}$ (derived in Appendix~\ref{appendix:IB_energy}) were evaluated independently from the steady-state solutions for the system at a given $\tilde{T}$ over the range $\tilde{T}=1$ to $5$. Applying $\tilde{\Gamma}_{xs}$ (Eq.~\ref{eq:inv_solute_xs}), the right side of the adsorption equation $-\tilde{\Gamma}_{xs}\frac{d\tilde{\mu}_e}{d\tilde{T}}$ is obtained; here, $\tilde{\mu}_e(\tilde{T})$ is determined from the bulk-phase coexistence.
The derivative term on left side of the adsorption equation $d\tilde{\gamma}/d\tilde{T}$ was evaluated numerically. The agreement between the two terms is shown in inset of Fig.~\ref{fig:fe_vs_T_Gibbs_ads}b (for 200 calculations between $\tilde{T} = 1$ and $5$) and validates the definition of $\tilde{\Gamma}_{xs}$ and the agreement of the phase-field formulation with the Gibbs adsorption relation.

\begin{figure}[h!]
    \centering
    \begin{subfigure}[t]{0.5\textwidth}
        \centering
        \includegraphics[width=1\textwidth]{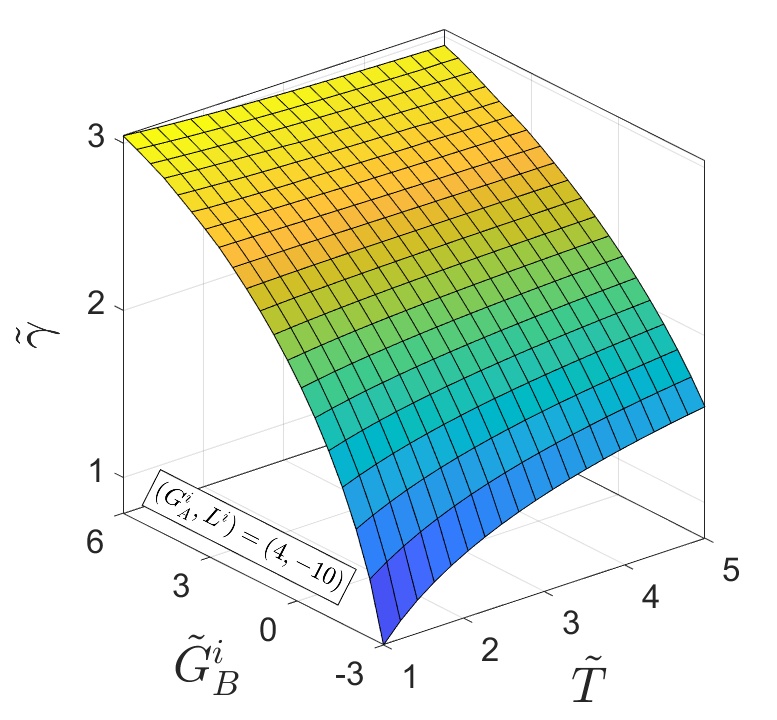}
        % \caption{Lorem ipsum}
    \end{subfigure}%
    ~ ~    
    \begin{subfigure}[t]{0.5\textwidth}
        \centering
        \includegraphics[width=1\textwidth]{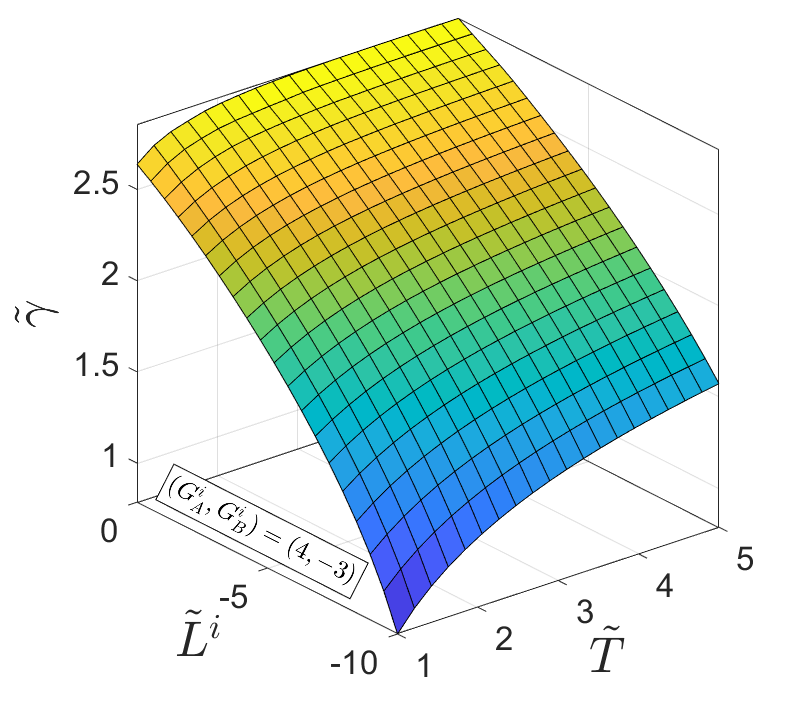}
        % \caption{Lorem ipsum}
    \end{subfigure}%
    \caption{Surface plots showing the dependence of IB energy $\tilde{\gamma}$ on: (a) IB parameter $\tilde{G}^i_B$ and temperature $\tilde{T}$ for $(\tilde{G}^i_A,\tilde{L}^i) = (4,-10)$, i.e. IB with favorable $A$-$B$ mixing energetics; (b) IB interaction parameter $\tilde{L}^i$ and $\tilde{T}$ for $(\tilde{G}^i_A,\tilde{G}^i_B) = (4,-3)$, i.e. IB with favorable energetics for pure $B$ over pure $A$. $\tilde{\varepsilon}^2 = 3$.} 
    \label{fig:gamma_surf_GiB_T}
\end{figure}

Surface plots depicting the variation of $\tilde{\gamma}$ with $\tilde{T}$ and an IB parameter ($\tilde{G}^i_B$ or $\tilde{L}^i$) are shown in Fig.~\ref{fig:gamma_surf_GiB_T}. Low $\tilde{\gamma}$ is obtained at low $\tilde{T}$ and increasingly negative $\tilde{L}^i$ (i.e. highly favorable $A$--$B$ interaction) or $\tilde{G}^i_B-\tilde{G}^i_A$ (i.e. low pure component $B$ energy relative to $A$). Variation in $\tilde{\gamma}$ with $\tilde{T}$ is negligible for large $\tilde{G}^i_B$; this is due to $\tilde{\Gamma}_{xs} \rightarrow 0$. For low $\tilde{G}^i_B$ or $\tilde{L}^i$, the change in $\tilde{\gamma}$ with $\tilde{T}$ is significant since $\tilde{\Gamma}_{xs}$ is positive and increases with decreasing $\tilde{G}^i_B$ or $\tilde{L}^i$. The IB parameter regime corresponding to low $\tilde{G}^i_B$ and $\tilde{L}^i$ is potentially most useful for design purposes at given $\tilde{T}$ as it corresponds to the lowest $\tilde{\gamma}$ despite a significant increase in $\tilde{\gamma}$.

\FloatBarrier

\begin{figure}[h!]
    \centering
    \begin{subfigure}[t]{0.5\textwidth}
        \centering
        \includegraphics[width=1\textwidth]{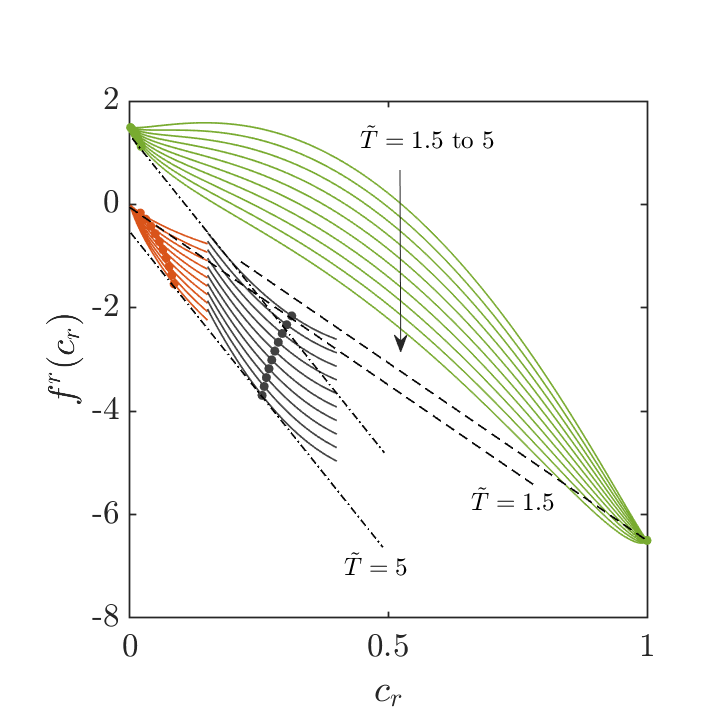}
    %     % \caption{Lorem ipsum}
    \end{subfigure}%
    ~ ~    
    \begin{subfigure}[t]{0.5\textwidth}
        \centering
        \includegraphics[width=1\textwidth]{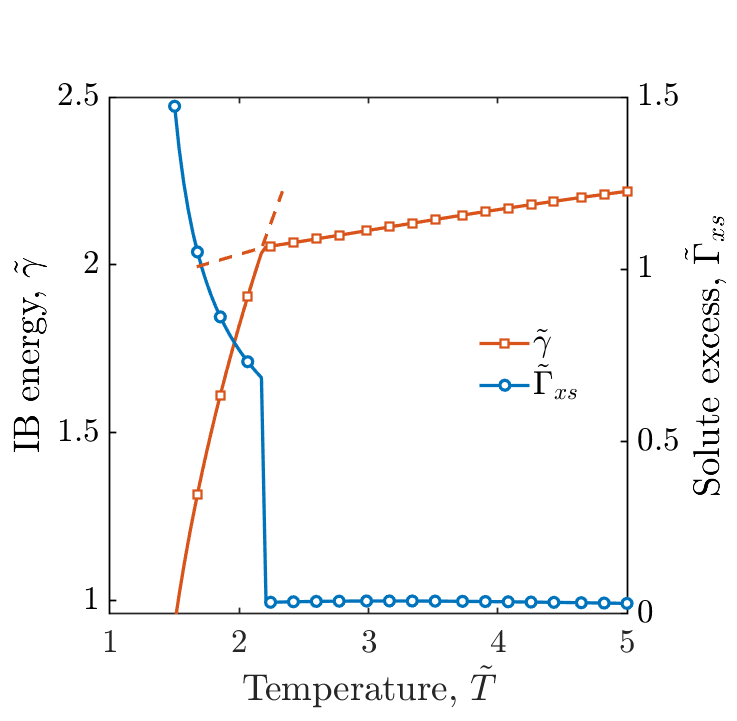}
        % \caption{Lorem ipsum}
    \end{subfigure}%
    \caption{(a) Free energy density $f^r$ versus phase concentration $c_r$ curves for $r=m,i,p$. The IB free energy curves $f^i$ plotted for $(\tilde{G}^i_A,\tilde{G}^i_B,\tilde{L}^i)=(1.5,-6.5,15)$ show saddle point due to $L^i>>0$. (b) Variation in IB energy $\tilde{\gamma}$ and solute excess $\tilde{\Gamma}_{xs}$ with $\tilde{T}$ for the system defined in (a) demonstrates first-order IB phase transition.} 
    \label{fig:adsorption_IB_spinodal}
\end{figure}

Free energy curves at different temperatures $(5 > \tilde{T} > 1.5)$ for an IB phase given by $(\tilde{G}^i_A,\tilde{G}^i_B,\tilde{L}^i)=(1.5,-6.5,15)$ are shown in Fig.~\ref{fig:adsorption_IB_spinodal}a. As the IB interaction parameter $\tilde{L}^i$ is chosen to be positive and large the IB free energy curves $\tilde{f}^i$ exhibit a saddle point. The variation in the corresponding excess properties are shown in Fig.~\ref{fig:adsorption_IB_spinodal}b. The hypothetical IB phase exhibits a first-order transition with respect to $\tilde{T}$, as demonstrated by the discontinuity in $\tilde{\Gamma}_{xs}$ and the change in slope of $\tilde{\gamma}$. At the critical phase transition temperature, the two different IB states exhibit the same $\tilde{\gamma}$. At low $\tilde{T}$'s, $\tilde{\gamma}$ increases sharply corresponding to a large and positive $\tilde{\Gamma}_{xs}$ (following Eq.~\ref{eq_kks:gibbs_ads}). Beyond the transition point the dependence of $\tilde{\gamma}$ on $\tilde{T}$ is significantly lower as $\tilde{\Gamma}_{xs}$ is small. Note that lower magnitudes of $\tilde{G}^i_A$ and $\tilde{G}^i_B$ are chosen to be obtain ${\gamma}$ ($\approx 0.36 \tilde{\gamma}$ J/m$^2$) values within $1$ J/m$^2$.

% % \FloatBarrier
% \begin{figure}[h!]
%     \centering
%     \begin{subfigure}[t]{0.5\textwidth}
%         \centering
%         \includegraphics[width=1\textwidth]{kks_figures/excess_surface_GiB_T.png}
%         % \caption{Lorem ipsum}
%     \end{subfigure}%
%     ~ ~    
%     \begin{subfigure}[t]{0.5\textwidth}
%         \centering
%         \includegraphics[width=1\textwidth]{kks_figures/excess_surface_LiAB_T.png}
%         % \caption{Lorem ipsum}
%     \end{subfigure}%
%     \caption{} 
%     \label{fig:gamma_surf_LiAB_T}
% \end{figure} 

% Hypothetical first-order IB transition. (a) Free energy density versus concentration curves for a bulk system described in Fig.~\ref{fig:adsorption_IB_spinodal}(a) and IB phase $(\tilde{G}^i_A,\tilde{G}^i_B,\tilde{L}^i)=(2,-6,15)$ for temperatures $\tilde{T}=1.5$ to $5$. Filled marker on an IB free energy curve represents the IB state in equilibrium with the bulk states (markers on $f^m$ and $f^p$ curves) at the corresponding $\tilde{T}$. (b)

% \newpage
\FloatBarrier
\section{Discussion} \label{sec:discussion}

\subsection{Phase-field model}

In our phase-field description (Sec.\,\ref{subsec:pf_model}), we introduced an IB phase that can be assigned a distinct free energy $f^i$. This follows in line with the treatment of free surfaces and grain boundaries in classical thermodynamic models \cite{johnson1979interfacial,lejcek2010grain,frolov2015phases}, and in phase-field models for GB segregation where the GB is implicitly or explicitly described by a distinct free energy \cite{cha2002phase,abdeljawad2015stabilization,kim2016GBsegregation}. Piecewise interpolating functions are used to effectively define the interface to be composed of: a mixture of matrix and IB phase ($0<\phi<0.5$), an IB phase at the interface center ($\phi=0.5$), and a mixture of IB and precipitate phases ($0.5<\phi<1$). (Note that $\phi=0.5$ is chosen to define $f^i(c_i)$ for analytic convenience; the diffuse IB region is defined over $0<\phi<1.0$.) On the other hand, conventional two-phase models describe the diffuse IB purely as a mixture of the bulk matrix and precipitate phases. Following customary KKS approach \cite{kim1999phase,cha2002phase,kim2016GBsegregation}, we introduce phase concentrations as model variables, and at any infinitesimal element, we impose the condition (Eq.\,\ref{eq_kks:equal_potential}) of equal chemical potential ${\mu}$ between the hypothetical phases that make up the point. Thereby, our model inherits the following advantages \cite{kim1999phase,hu2007thermodynamic,provatas2011phase} of the KKS formulations: (i) free energies with large curvatures can be employed (as in the schematic Fig.\,\ref{fig:interpol_fe_schemetic}), which is especially useful to model an intermetallic precipitate; (ii) for given interpolating functions, steady-state IB properties arise solely from the equilibrium states of the matrix, IB and precipitate phases; (iii) steady-state solutions are obtained analytically.

At steady state, the standard conditions for equilibrium between between bulk phases, given by the common tangent construction, and for equilibrium between a bulk and interface phase, given by the parallel tangent construction \cite{guggenheim1985thermodynamics,fowler1941statistical,mcleangrain,johnson1979interfacial,howe1997interfaces,lejcek2010grain}, are recovered (Eqs.\,\ref{eq_kks:equilibrium_potential},\ref{eq_kks:pt_eqn_We_def} and \ref{eq_kks:ct_eqn}). The equilibrium states of the phases govern the properties across the system. The parallel tangent distance $f^i(c^e_i) - f^m(c^e_m) - (c^e_i-c^e_m)\mu_e \equiv W_e$ defines the excess grand potential across interface via $\Omega(x) = 2W_e I(\phi_e)$. Here, $W_e$ replaces the fitting parameter $W_\phi$ in the conventional KKS model, and $I$ serves as the double-well barrier potential, in addition to serving as an interpolating function.
The present formulation is analogous to the GB segregation model developed by Cha et al. \cite{cha2002phase} which introduces a distinct free energy and phase concentration for the GB phase and imposes the equal chemical potential condition between the matrix and the GB. Therefore, the steady-state properties (Sec.\,\ref{sec:param_study:steady-state}) associated with the phase-field $\phi_e$, via the parallel tangent distance $W_e$, i.e. IB energy $\gamma$, IB width $\lambda$ obtained in the present work are similar in form to that obtained for GB segregation in \cite{cha2002phase}. The differences arise from the fact that 
the chemical identities of the adjoining phases in the present case of an IB are distinct ($f^p\not\equiv f^m$). 

We can further relate the results obtained in our phase-field formulation with results from classical non-gradient treatments. In the non-gradient treatment of the IB phase, such as in a Guggenheim model \cite{guggenheim1985thermodynamics}, the system is composed of only the homogeneous phases $m$, $i$ and $p$, separated by two sharp interfaces. If the thickness $\delta$ and the number of atoms of the IB layer are fixed, then only exchange of atoms with the IB is allowed. Therefore, the equilibrium state for the IB is given by the parallel tangent construction and the distance by $W_e$ \cite{hillert1975lectures}. We may adapt to the IB, the definition of the parallel tangent distance for the GB \cite{hillert1975lectures}. Thus, $W_e$ represents the increase in free energy if a unit volume of a new IB phase with concentration $c^e_i$ is created from the $m$ phase with concentration $c^e_m$ (or equivalently from the $p$ phase with concentration $c^e_p$) in a system with large bulk phases. The IB energy in this non-gradient model is given by $\gamma = W_e\delta$, which is identical in form to the IB energy obtained in our model, $\gamma \approx 1.2 W_e \lambda$ (Eq.\,\ref{eq:gamma_int}). While the IB thickness $\delta$ is fixed in the non-gradient model, and therefore independent of the IB state ($c^e_i$), in the phase-field model the width is a function of the IB state and concentration $c^e_i$ via $\lambda \propto \varepsilon/\sqrt{W_e}$. In the present work, the gradient energy coefficient $\varepsilon$ is the fixed quantity and therefore $\gamma \propto \varepsilon \sqrt{W_e}$. A similar difference arises between the non-gradient and gradient descriptions of GB segregation, which has been discussed by S. G. Kim et al. \cite{kim2016GBsegregation}.

\subsection{Solution thermodynamics} \label{sec:disc_soln_behv}

In Sec.\,\ref{sec:regular_soln}, we assumed the regular solution behavior, and performed a parametric study to demonstrate the relationship between the IB-phase parameters $(G^i_A,G^i_B,L^i)$ and steady-state properties of the diffuse-interface model. For a given system and temperature, the parallel tangent condition yields a segregation equation (Eq.\,\ref{eq:cie_parallel_tgt}) that determines the state $c^e_i$ of the IB phase from the two-phase equilibrium given by $\mu_e$. While this form is appropriate for the binary system, Eq.\,\ref{eq:cie_parallel_tgt} may be recast by substituting $\mu_e$ with the derivative $df^m(c^e_m)/dc^e_m$ evaluated using the regular solution model for $f^m$. This yields a relation of the form $c^e_i/(1-c^e_i) = c^e_m/(1-c^e_m)\exp{\left(\Delta E_{seg}v_m/RT\right)}$, which is identical to the GB segregation isotherms of Fowler-Guggenheim \cite{fowler1941statistical,lejcek2010grain} for non-ideal interactions, and the Langmuir-McLean isotherm \cite{mcleangrain,lejcek2010grain} in the ideal solution limit ($L^m=L^i=0$); here, $\Delta {E}_{seg}$ is the segregation energy, which is determined by $(G^i_A,G^i_B,L^i)$ for a given bulk system. Using this form, it could be possible to obtain the IB phase parameters by matching the segregation energy with that obtained from first principles calculations \cite{shin2017solute}.
Traditionally, the GB segregation isotherms are used to evaluate the dependence of GB concentration $c^e_i$ on the grain concentration $c^e_m$; where, $c^e_m$ can be varied by controlling the total concentration in the system. However, for the two-phase system, $c^e_m$ is fixed with respect to total concentration by the common tangent construction. Therefore, $c^e_m$ can be varied only as a function of $T$ through a change in the bulk phase equilibrium $\mu_e$, as demonstrated in Sec.\,\ref{subsec:excess_GA}. This situation is similar to that for GB segregation in the presence of a second phase in the system~\cite{kaplan2013review}.

Free surfaces and GBs are known to exhibit phase-like behavior, including spinodal decomposition \cite{johnson1979interfacial,howe1997interfaces,kaplan2013review,da2019thermodynamics}. However, the IB has received little experimental or theoretical attention in this context. Using the model developed in this work, attractive solute-solvent interactions ($L^i<0$) were used in Sec\,\ref{sec:regular_soln} to model solution behavior of IB and repulsive interactions ($L^i>0$) were used to demonstrate first-order transition at IB (Fig.\,\ref{fig:adsorption_IB_spinodal}). However, limitations of the solution models apply; for instance, while $L^i<<0$ will produce ordering in physical systems, the configurational entropy assumes random mixing. Additionally, the IB phase parameters $(G^i_A,G^i_B)$ are not well defined physical quantities. In the context of GBs, the difference between the pure component energies of the GB and grain ($G^i_{A/B}-G^m_{A/B}$) represents the increase in energy if a unit volume of new boundary is created in a system of pure $A/B$ of the $m$ phase and can be directly related to GB energy of $A/B$ \cite{hillert1975lectures}. The creation of the IB, however, cannot be realized in a single component phase, and therefore cannot be directly related to experimental measurement. Nevertheless, the equilibrium state of the IB is itself well defined since it is associated with measurable excess quantities ($\gamma$,$\Gamma_{xs}$).

\subsection{Gibbs adsorption}

The IB phase concentration $c^e_i$ is useful for statistical thermodynamic description of IB segregation as discussed in the previous section. However, the excess solute concentration is the relevant classical thermodynamic quantity that is independent of model assumptions \cite{cahn1979interfacial}.
In Sec\,\ref{subsec:excess_GA} we adapted the definition of solute excess $\Gamma_{xs}$ proposed for two-component diffuse-interface models by McFadden and Wheeler \cite{mcfadden2002gibbs}; this definition is independent of the Gibbs dividing surface convention. Using this form, we showed that our model satisfies the Gibbs adsorption equation $\frac{d\gamma}{dT}=-\Gamma_{xs}\frac{d\mu_e}{dT}$. 
 The Gibbs adsorption behavior, given by the variation of $\gamma$ and $\Gamma_{xs}$ with $T$, was demonstrated using ideal solution behavior of the IB phase. Under the regular solution behavior, different IB phases $(G^i_A,G^i_B,L^i)$ are expected to exhibit distinct Gibbs adsorption behaviors. Therefore, for a given bulk system, $f^i$ or $(G^i_A,G^i_B,L^i)$ could be chosen to match the Gibbs adsorption behavior obtained experimentally or using atomistic simulations. In this regard, indirect measures of interfacial energy (such as based on coarsening kinetics \cite{ardell1966coarsening,howe1997interfaces} or atomistic simulations \cite{hyland1992homogeneous,howe1997interfaces}) are possible. In multicomponent alloys, experimental determination of the relative solute excess from atom probe tomography \cite{krakauer1993absolute, marquis2006composition}) is possible and has been used to quantify segregation.

% \subsection{Future directions}
In the present study, the phase-field formulation was presented for the simplest case of IB segregation in a binary component alloy. Here, $T$ is the only degree of freedom available to effect a change in the equilibrium state of the bulk phases, and therefore the IB phase. However, the ternary system is expected to possess a compositional degree of freedom in that the IB state $(\gamma,\Gamma_{xs})$ can be varied with system composition, in addition to $T$. For example, the Cahn-Hilliard formulation by Dregia and Wynblatt \cite{dregia1991equilibrium}) showed a variation in adsorption of Au at the interface between Al-rich and Cu-rich phases with variation in the total Au composition in the phase-separating Al-Cu system.
In precipitation hardening systems, it is the microalloying elements that are often found to segregate at the IB \cite{biswas2010simultaneous,biswas2011precipitates,marquis2003mg,liu2016interphase}. The framework proposed in this paper can be extended to multicomponent systems, and therefore should allow for direct comparison with experimental adsorption behaviors.

% \subsection{Numerical Implementation}
% In the model presented here, we used (Sec\,\ref{subsec:pf_model}) piecewise interpolating functions that are $C^1$-continuous for $\phi \in [0,1]$. The commonly used double-well potential $16\phi^2(1-\phi)^2$ satisfied $C^1$-continuity and hence could be utilized for the interpolating functions. For simulating microstructure evolution, for example using finite element discretization and the diffusive form of composition evolution, numerical implementation will require $C^2$-continuous interpolating functions (\textcolor{red}{thats only tru if you solvee the full Cahn-Hilliard, but an operator split does not require that. I think MOOSE allows both approaches}). A possible choice for such an implementation is given below for $-1 < \phi < 1$ describing the interface, and $\phi = -1$ and $\phi = 1$ denoting $m$ and $p$ phases. 
% \begin{flalign} \nonumber
% % I(\phi) = \begin{cases}
%  &\mathrm{For} \text{ $\phi \leq 0$}: \quad I(\phi) = 1-\frac{1}{2}(3\phi^5-5\phi^3), \quad M(\phi) = 1-I(\phi), \quad P(\phi) = 0 \\
%  &\mathrm{For} \text{ $\phi > 0$}: \quad I(\phi) = 1+\frac{1}{2}(3\phi^5-5\phi^3), \quad M(\phi) = 0, \qquad \qquad P(\phi) = 1-I(\phi) &&
% %  \end{cases}
% \end{flalign}
% where $M$, $P$ and $I$ are interpolating functions corresponding to the $m$, $p$ and $i$ phases, as before. Using this, our model can be readily implemented in the MOOSE framework \cite{schwen2017rapid,aagesen2017quantifying,jokisaari2017benchmark}, which is an open-source FEM-based solver for phase-field simulations.

\FloatBarrier
\section{Conclusions}

In this study, we presented a diffuse interface approach for interphase boundary (IB) segregation by defining a compositionally-homogeneous IB phase with energetics and composition-dependence independent from the adjoining bulk phases. The interface between the two bulk phases in this description consists of the IB phase and its gradients with the adjoining bulk phases; this distinguishes the present model from traditional diffuse interface models that describe the interface as a gradient region between the bulk phases. This description allows the interfacial properties to be described from the classical parallel tangent condition for equilibrium between an interface-phase and the bulk phase. This also allows incorporation of diverse free energy-composition dependencies for the IB phase. We also showed that excess properties--interfacial energy and solute excess-- corresponding to the Gibbsian (2D) interface can be derived from the diffuse interface formulation, consistent with the Gibbs adsorption equation. The excess properties and the adsorption equation are of practical importance as they will allow comparison with measurements from experiments and atomistic simulations. Further work using the formulation is expected with regards to its numerical implementation, extension to multicomponent systems and coupling with elasticity models. The current formulation therefore provides a potential tool to simulate the effect of IB segregation on the thermal stability of precipitates against coarsening, such as those observed in Al-Sc-Mg and Mg-Sn-Zn.
\appendix

\renewcommand{\thesection}{\Alph{section}} % \Alph{section}
\renewcommand{\theequation}{\Alph{section}\arabic{equation}}
\section{Steady-state phase field profile} \label{appendix:profile}

Here we derive the equation governing the steady-state phase field $\phi_e(x)$ in a one-dimensional system. Regarding the phase concentrations $c_r$ ($r=m,i,p$) as functions of $c$ and $\phi$, and differentiating $c(x)$ defined in Eq.\,\ref{eq_kks:effective_conc_variable}, we get
\begin{flalign} \label{aeq_kks:diff_conc_c}
    M(\phi) \frac{\partial c_m}{\partial c} + I(\phi) \frac{\partial c_i}{\partial c} +P(\phi) \frac{\partial c_p}{\partial c} = 1 &&
\end{flalign}
and
\begin{flalign} \label{aeq_kks:diff_conc_phi}
    M(\phi) \frac{\partial c_m}{\partial \phi} + I(\phi) \frac{\partial c_i}{\partial \phi} + P(\phi) \frac{\partial c_p}{\partial \phi} = \frac{d M(\phi)}{d \phi}\,(c_i-c_m) + \frac{d P(\phi)}{d \phi}\,(c_i-c_p) &&
\end{flalign}

Relations \ref{aeq_kks:diff_conc_c} and \ref{aeq_kks:diff_conc_phi} can be used to obtain the derivatives of $f(c,\phi)$ (Eq.\,\ref{eq_kks:f_local}) as
\begin{flalign} \label{aeq_kks:non-eq_dfdc} \nonumber
    \frac{\partial f}{\partial c} &= M(\phi) \frac{d f^m(c_m)}{d c_m}\frac{\partial c_m}{\partial c}  + I(\phi) \frac{d f^i(c_i)}{d c_i}\frac{\partial c_i}{\partial c} + P(\phi) \frac{d f^p(c_p)}{d c_p}\frac{\partial c_p}{\partial c} \\
    &= \left(M(\phi)\frac{\partial c_m}{\partial c} + I(\phi)\frac{\partial c_i}{\partial c} + P(\phi)\frac{\partial c_p}{\partial c} \right)\mu = \mu &&
\end{flalign}
and
\begin{flalign} \label{aeq_kks:non-eq_dfdphi} \nonumber
    \frac{\partial f}{\partial \phi} =& M(\phi)\frac{df^m(c_m)}{dc_m}\frac{\partial c_m}{\partial \phi} + I(\phi)\frac{df^i(c_i)}{dc_i}\frac{\partial c_i}{\partial \phi} + P(\phi)\frac{df^p(c_p)}{dc_p}\frac{\partial c_p}{\partial \phi}  \\ \nonumber
    &+ \frac{dM(\phi)}{d\phi} f^m(c_m)  + \frac{dI(\phi)}{d\phi} f^i(c_i) + \frac{dP(\phi)}{d\phi} f^p(c_p) \\ \nonumber
    =& \left(M(\phi)\frac{\partial c_m}{\partial \phi} + I(\phi)\frac{\partial c_i}{\partial \phi} + P(\phi)\frac{\partial c_p}{\partial \phi} \right)\mu \\ \nonumber
    &+ \frac{dM(\phi)}{d\phi} \left(f^m(c_m)-f^i(c_i)\right) + \frac{dP(\phi)}{d\phi} \left(f^p(c_p)-f^i(c_i)\right) \\
    =& \left(f^m(c_m) - f^i(c_i) - (c_m-c_i)\mu\right)\frac{dM(\phi)}{d\phi} + \left(f^p(c_p) - f^i(c_i) - (c_p-c_i)\mu\right)\frac{dP(\phi)}{d\phi} &&
\end{flalign} 
where the condition for equal chemical potential condition at a point $\mu(x)$ (Eq.\,\ref{eq_kks:equal_potential}) and the identity $dI/d\phi = -dM/d\phi - dP/d\phi$ were used.

For a one-dimensional system at steady state, the phase field $\phi_e(x)$ satisfies Eq.\,\ref{eq_kks:1D_eq_phi} given as
\begin{flalign} \label{aeq_kks:eq_phase-field_begin}
    \frac{\partial f}{\partial \phi_e} = \varepsilon^2 \frac{d^2\phi_e}{dx^2} && 
\end{flalign}
The phase concentrations $c_r(x)$ are constant across $x$ at steady state and are given by $c^e_r$. Using \ref{aeq_kks:non-eq_dfdphi} in \ref{aeq_kks:eq_phase-field_begin}, we get
\begin{flalign}
    -W^m_e\frac{dM(\phi_e)}{d\phi_e} - W^p_e\frac{dP(\phi_e)}{d\phi_e} = \varepsilon^2 \frac{d^2\phi_e}{dx^2} &&
\end{flalign}
where $W^m_e \equiv f^i(c_i^e) - f^m(c_m^e) - (c_i^e-c_m^e)\mu_e$ and $W^p_e \equiv f^i(c_i^e) - f^p(c_p^e) - (c_i^e-c_p^e)\mu_e$ are constant across $x$. Multiplying both sides by $d\phi_e/dx$, integrating from $x = -\infty$ to $x = +\infty$, and changing the variable of integration to $\phi_e$, we get
\begin{flalign}
    W^m_e \int_{0}^{0.5} \frac{dM(\phi_e)}{d\phi_e} d\phi_e + W^p_e \int_{0.5}^{1} \frac{dP(\phi_e)}{d\phi_e} d\phi_e = 0 &&
\end{flalign}
Therefore, $W^m_e = W^p_e \equiv W_e$ or $f^m(c_m^e)-f^p(c_p^e)-(c_m^e-c_p^e)\mu_e=0$, which along with the Eq.\,\ref{eq_kks:equilibrium_potential} defines the common tangent construction between $f^m$ and $f^p$, and $W_e$ defines the vertical distance between the common tangent and the parallel tangent to $f^i$.
Using $W^m_e = W^p_e \equiv W_e$ in (\ref{aeq_kks:eq_phase-field_begin}) and integrating after multiplying by $d\phi_e/dx$ gives
\begin{flalign} \label{aeq_kks:phase-field_profile_final}
    \frac{\varepsilon^2}{2}\left(\frac{d\phi_e}{dx}\right)^2 = W_e I(\phi_e) &&
\end{flalign}
where the identities $-dM/d\phi_e = dI/d\phi_e$ for $\phi_e \in [0, 0.5]$ and $-dP/d\phi_e = dI/d\phi_e$ for $\phi_e \in (0.5, 1]$ were used.

\section{Interphase boundary energy} \label{appendix:IB_energy}

Here we derive the interfacial excess properties by defining the Gibbs dividing surface at $x=0$. The solute excess $C_{xs}$, assuming equal molar volumes $v_m$ in each phase, is evaluated as \cite{wheeler1993phase}
\begin{flalign} \label{aeq_kks:solute_excess} \nonumber
    C_{xs} &= \int_{-l}^{+l} c_e(x) dx - \int_{-l}^0 c_m^e dx - \int_0^{+l} c_p^e dx \\ \nonumber
    % &= \int_{-l}^{0} \left(M(\phi_e)c_m^e + P(\phi_e)c_p^e + I(\phi_e)c_i^e - c_m^e \right)dx \\
    % &\qquad + \int_{0}^{+l} \left(M(\phi_e)c_m^e + P(\phi_e)c_p^e + I(\phi_e)c_i^e - c_p^e \right) dx \\
    &= \int_{-l}^{0} \left(M(\phi_e)c_m^e + \left(1-M(\phi_e)\right)c_i^e - c_m^e \right)dx  + \int_{0}^{+l} \left(P(\phi_e)c_p^e + \left(1-P(\phi_e)\right)c_i^e - c_p^e \right) dx \\ \nonumber
    % &= (c_i^e-c_m^e) \int_{-l}^{0} \left(1-M(\phi_e)\right) dx \\
    % &\qquad + (c_i^e-c_p^e) \int_{0}^{+l} \left(1-P(\phi_e)\right) dx
    % &= (c_i^e-c_m^e) \int_{-l}^{0} I(\phi_e) dx + (c_i^e-c_p^e) \int_{0}^{+l} I(\phi_e) dx \\ \nonumber
    &= (c_i^e-c_m^e) \int_{0}^{0.5} I(\phi_e) \frac{dx}{d\phi_e}d\phi_e + (c_i^e-c_p^e) \int_{0.5}^{1} I(\phi_e) \frac{dx}{d\phi_e}d\phi_e \\
    &= (2c_i^e - c_m^e - c_p^e) \frac{\varepsilon}{3\sqrt{2W_e}} &&
\end{flalign}

The interphase boundary (IB) energy $\gamma$, defined as the excess free energy of the system per unit area due to the presence of the IB, is defined as \cite{lupis1983chemical,wheeler1992phase,mcfadden2002gibbs,provatas2011phase}
\begin{flalign} \label{aeq_kks:IB_energy_def}
    \gamma = (F-F_{o}) - C_{xs} \mu_e &&
\end{flalign}
where $F$ is the total free energy (Eq.\,\ref{eq:F_functional}) of the diffuse-interface system and $F_{o}$ is the free energy of the reference system whose matrix and precipitate properties remain homogeneous up to the dividing surface at $x=0$. Therefore,
\begin{flalign} \label{aeq_kks:IB_energy_expand}
    \gamma =& \int_{-l}^{+l} \left[f(c_e,\phi_e) + \frac{\varepsilon^2}{2}\left(\frac{d\phi_e}{dx}\right)^2 \right] dx - \int_{-l}^{0} f^m(c_m^e) dx - \int_{0}^{+l} f^p(c_p^e) dx - C_{xs} \mu_e &&
    % &- \int_{-l}^{+l} \left[M(\phi_e)c_m^e + P(\phi_e)c_p^e + I(\phi_e)c_i^e \right]\mu_e dx \\
    % &+ \int_{-l}^{0} c_m^e dx + \int_{0}^{+l} c_p^e dx  &&
\end{flalign}
Substituting for $f$ (Eq.\,\ref{eq_kks:f_local}) and $C_{xs}$ (\ref{aeq_kks:solute_excess}), and reorganizing the terms to evaluate the integrals piecewise over $[-l,0]$ and $(0,+l]$, we get
\begin{flalign} \label{aeq_kks:IB_energy_piecewise_x}
    \gamma =& \int_{-l}^{+l} \frac{\varepsilon^2}{2}\left(\frac{d\phi_e}{dx}\right)^2 dx \\ \nonumber
    &+ \int_{-l}^{0} \left[{M(\phi_e)\left(f^m(c_m^e) - c_m^e \mu_e \right) +\left(1-M(\phi_e)\right) \left(f^i(c_i^e) - c_i^e \mu_e \right) - f^m(c_m^e) - c_m^e \mu_e } \right] dx \\ \nonumber
    &+ \int_{-l}^{0} \left[{P(\phi_e)\left(f^p(c_p^e) - c_p^e \mu_e \right) +\left(1-P(\phi_e)\right) \left(f^i(c_i^e) - c_i^e \mu_e \right) - f^p(c_p^e) - c_p^e \mu_e } \right] dx &&
\end{flalign}
where the identities $I(\phi_e(x)) = 1-M(\phi_e(x))$ on $[-l,0]$ and $I(\phi_e(x)) = 1-P(\phi_e(x))$ on $(0,+l]$ were used. Simplifying the second and third terms,
\begin{flalign} \nonumber
    \gamma =& \int_{-l}^{+l} \frac{\varepsilon^2}{2}\left(\frac{d\phi_e}{dx}\right)^2 dx + \int_{-l}^{+l} W_e I(\phi_e) dx \\
    =& \int_{-l}^{+l} \varepsilon^2 \left(\frac{d\phi_e}{dx}\right)^2 dx = \int_0^1 \varepsilon^2 \frac{d\phi_e}{dx} d\phi_e && \label{aeq_kks:IB_energy_gradient}
\end{flalign}
where the definition of $W_e$ (Eq.\,\ref{eq_kks:pt_eqn_We_def}) and the equality \ref{aeq_kks:phase-field_profile_final} were used.

\section{Gibbs adsorption} \label{appendix:GA}

Here, we adapt to our phase-field model, the approach presented by McFadden and Wheeler \cite{mcfadden2002gibbs} to obtain the invariant solute excess $\Gamma_{xs}$ and the Gibbs adsorption equation for general diffuse interface models. We denote the system size $[-l,l]$ dependent properties using the subscript $l$. The free energy functional $F_l$ at steady state is given from Eq.\,\ref{eq:F_functional} as
\begin{flalign} \label{aeq:functional}
    F_l =& A_i \int_{-l}^{l} \left[f(c_e,\phi_e;T)+\frac{\varepsilon^2}{2}\left(\frac{d\phi_e}{dx}\right)^2 \right] dx &&
\end{flalign}  
The variation with respect to $T$ can be obtained as follows using the chain rule of differentiation and integration by parts to evaluate terms in the integrand
% \begin{flalign} \nonumber
%     \frac{dF_l}{dT} =& A_i \int_{-l}^{l} \left[\frac{\partial f}{\partial c_e}\frac{d c_e}{d T} + \frac{\partial f}{\partial T} + \frac{\partial f}{\partial \phi_e}\frac{d \phi_e}{d T} + \varepsilon^2 \frac{d\phi_e}{dx}  \cdot \frac{d^2\phi_e}{dxdT} \right] dx \\ \nonumber
%     =& A_i \int_{-l}^{l} \left[\mu_e \frac{dc_e}{dT} - s + \left(\frac{\partial f}{\partial \phi_e} - \varepsilon^2 \frac{d^2\phi_e}{d x^2} \right)\frac{d\phi_e}{dT} \right] dx \\
%     =& \mu_e\frac{dC_l}{dT} - S_l && \label{aeq:dFdT_form1} 
% \end{flalign}
\begin{flalign} \nonumber
    \frac{dF_l}{dT} =& A_i \int_{-l}^{l} \left[\frac{\partial f}{\partial c_e}\frac{d c_e}{d T} + \frac{\partial f}{\partial T} + \frac{\partial f}{\partial \phi_e}\frac{d \phi_e}{d T} + \varepsilon^2 \frac{d\phi_e}{dx}  \cdot \frac{d^2\phi_e}{dxdT} \right] dx \\ \nonumber
    =& \left. \frac{d\phi_e}{dT}\frac{d\phi_e}{dx} \right\vert_{-l}^{l} + A_i \int_{-l}^{l} \left[\mu_e \frac{dc_e}{dT} - s + \frac{\partial f}{\partial \phi_e}\frac{d\phi_e}{dT} - \varepsilon^2 \frac{d\phi_e}{dT}  \cdot \frac{d^2\phi_e}{dx^2} \right] dx \\ \nonumber
        =& A_i \int_{-l}^{l} \left[\mu_e \frac{dc_e}{dT} - s + \left(\frac{\partial f}{\partial \phi_e} - \varepsilon^2 \frac{d^2\phi_e}{d x^2} \right)\frac{d\phi_e}{dT} \right] dx \\
    =& \mu_e\frac{dC_l}{dT} - S_l && \label{aeq:dFdT_form1} 
\end{flalign} 
% \begin{flalign}
%     \int_{-l}^{l} \varepsilon^2 \frac{d\phi_e}{dx}  \cdot \frac{d^2\phi_e}{dxdT} dx = \left. \frac{d\phi_e}{dT}\frac{d\phi_e}{dx} \right\vert_{-l}^{l} - \int_{-l}^{l} \varepsilon^2 \frac{d\phi_e}{dT}  \cdot \frac{d^2\phi_e}{dx^2} dx
% \end{flalign} 
where \ref{aeq_kks:eq_phase-field_begin} was used. The boundary term in the second line is eliminated using $d\phi_e/dx \rightarrow 0$ for $\pm l$ far from the IB. $s=s(c_e)$ is the local (configurational) entropy density. \ref{aeq:functional} can also be written as
\begin{flalign}
    F_l =& A_i \int_{-l}^{l} \left[\frac{\varepsilon^2}{2}\left(\frac{d\phi_e}{dx}\right)^2 + \left(f(c_e,\phi_e,T)-f^m(c^e_m,T)-\mu_e(c_e-c^e_m) \right) \right] dx \nonumber \\ \nonumber
    &\, - \left(c^e_m\mu_e - f^m(c^e_m,T)\right)A_i(2l) + \mu_e A_i \int_{-l}^{l} c_e \, dx \\
    =& A_i \gamma_l - PV_l + \mu_eC_{l} &&
\end{flalign}
where $P$ is the equilibrium pressure in the bulk phases; $\gamma_l$, $V_l$ and $C_l$ are the IB energy, volume and total concentration of the system $[-l,l]$. 
% For $l \rightarrow \infty$, $\gamma_l$ converges to $\gamma$. 
Taking derivative with respect to $T$ gives
\begin{flalign} \label{aeq:dFdT_form2}
    \frac{dF_l}{dT} = A_i \frac{d\gamma_l}{dT} - \frac{dP}{dT}V_l + \mu_e \frac{dC_l}{dT} + \frac{d\mu_e}{dT}C_l &&
\end{flalign}
Subtracting \ref{aeq:dFdT_form1} from \ref{aeq:dFdT_form2} yields a version of the Gibbs adsorption equation as
\begin{flalign} \label{aeq:size_dep_GA}
    A_i\frac{d\gamma}{dT} = -\frac{d\mu_e}{dT}C_l - S_l + \frac{dP}{dT}V_l &&
\end{flalign}
Here, the integral quantities depend on choice of $l$ and $x=0$, except $\gamma_l$ which converges to $\gamma$ as $l \rightarrow \infty$ due to the common tangent construction. For the homogeneous bulk phases $m$ and $p$, the Gibbs-Duhem equations can be obtained in terms of the densities as
\begin{flalign} \label{aeq:GD_mat}
    0 &= -\frac{d\mu_e}{dT}c^e_m - s^m(c^e_m) + \frac{dP}{dT} \\ \label{aeq:GD_ppt}
    0 &= - \frac{d\mu_e}{dT}c^e_p - s^p(c^e_p) + \frac{dP}{dT} &&
\end{flalign}
Equations \ref{aeq:size_dep_GA}, \ref{aeq:GD_mat} and \ref{aeq:GD_ppt} can be written in the matrix form as
\begin{flalign}
    \begin{pmatrix}
        1 & V_l & -C_l \\
        0 & 1 & -c^e_m \\
        0 & 1 & -c^e_p
    \end{pmatrix}
    \begin{pmatrix}
        -A_i \frac{d\gamma}{dT} \\ \frac{dP}{dT} \\ 0
    \end{pmatrix}
    =
    \begin{pmatrix}
        S_l \\ s^m(c^e_m) \\ s^p(c^e_p)   
    \end{pmatrix} &&
\end{flalign}
Using Cramer's rule, $\frac{d\gamma}{dT}$ can be obtained. Expressing $V_l$, $C_l$ and $S_l$ in the integral form and reorganizing the terms yields
\begin{flalign} \label{aeq:GA_integral}
    \frac{d\gamma}{dT} &= -\int_{-l}^{l} \left[(s_e(x)-s^m(c^e_m))  - \frac{(s^p(c^e_p)-s^m(c^e_m))}{(c^e_p-c^e_m)}(c_e(x)-c^e_m) \right] dx &&
\end{flalign}
which can be recast as
\begin{flalign} \label{aeq:inv_solute_xs}
    \frac{d\gamma}{dT} = -\frac{d\mu_e}{dT} \int_{-l}^{l} \left[(c_e(x)-c^e_m)  - \frac{(c^e_p-c^e_m)}{(s^p(c^e_p)-s^m(c^e_m))}(s_e(x)-s^m(c^e_m)) \right] dx \equiv -\Gamma_{xs} \frac{d\mu_e}{dT} &&
\end{flalign}
This is an invariant form of the Gibbs adsorption equation derived following \cite{mcfadden2002gibbs} and \cite{umantsev2001continuum}. $\Gamma_{xs}$ is an invariant solute excess as defined by \cite{mcfadden2002gibbs}. Representing $c_e(x)$ and $s_e(x)$ using the interpolation scheme (as in Eq.\,\ref{eq:conc_profile}), the integral form of $\Gamma_{xs}$ in \ref{aeq:inv_solute_xs} can be expressed analytically as
\begin{flalign} \label{eq:anal_inv_excess}
    \Gamma_{xs} = C_{xs} - \frac{(c^e_p-c^e_m)}{(s^p(c^e_p)-s^m(c^e_m))}S_{xs} &&
\end{flalign}
where $S_{xs} = (2s^i(c^e_i)-s^m(c^e_m)-s^p(c^e_p))\varepsilon/(3\sqrt{2W_e})$ is the excess entropy is obtained with respect to the dividing surface at $x=0$ (analogous to $C_{xs}$ defined in \ref{aeq_kks:solute_excess}). While the definitions of $C_{xs}$ and $S_{xs}$ invoke the dividing surface $x=0$, $\Gamma_{xs}$ is itself independent of the choice. \ref{eq:anal_inv_excess} is specific to the present phase-field formulation.

\section*{Acknowledgement}

SP and SBK acknowledge support by the US National Science Foundation under Contract DMR-1554270. FA acknowledges start-up funds from Clemson University.

% \input{appendix_interface_thermodynamics.tex}

% KKS model
% \input{kks_theoretical_framework.tex}
% \pagebreak
% \afterpage{\blankpage}

%%% WBM model
% \input{wbm_model/theoretical_framework.tex}
% \FloatBarrier

% \input{wbm_model/equilibrium_properties.tex}

% \input{wbm_model/results_and_discussion.tex}
% \FloatBarrier

% \appendix
% \input{wbm_model/appendix.tex}
%%%

% \input{Conclusions}

% \section{ACKNOWLEDGMENTS}
% This work was primarily supported by the U.S. National Science Foundation under Grant No. DMR-1554270, with partial support from North Carolina State University.

\newpage

% \section*{References}
% \putbib
% \bibliographystyle{unsrt}
\bibliography{refs}

\begin{thebibliography}{65}%
\makeatletter
\providecommand \@ifxundefined [1]{%
 \@ifx{#1\undefined}
}%
\providecommand \@ifnum [1]{%
 \ifnum #1\expandafter \@firstoftwo
 \else \expandafter \@secondoftwo
 \fi
}%
\providecommand \@ifx [1]{%
 \ifx #1\expandafter \@firstoftwo
 \else \expandafter \@secondoftwo
 \fi
}%
\providecommand \natexlab [1]{#1}%
\providecommand \enquote  [1]{``#1''}%
\providecommand \bibnamefont  [1]{#1}%
\providecommand \bibfnamefont [1]{#1}%
\providecommand \citenamefont [1]{#1}%
\providecommand \href@noop [0]{\@secondoftwo}%
\providecommand \href [0]{\begingroup \@sanitize@url \@href}%
\providecommand \@href[1]{\@@startlink{#1}\@@href}%
\providecommand \@@href[1]{\endgroup#1\@@endlink}%
\providecommand \@sanitize@url [0]{\catcode `\\12\catcode `\$12\catcode
  `\&12\catcode `\#12\catcode `\^12\catcode `\_12\catcode `\%12\relax}%
\providecommand \@@startlink[1]{}%
\providecommand \@@endlink[0]{}%
\providecommand \url  [0]{\begingroup\@sanitize@url \@url }%
\providecommand \@url [1]{\endgroup\@href {#1}{\urlprefix }}%
\providecommand \urlprefix  [0]{URL }%
\providecommand \Eprint [0]{\href }%
\providecommand \doibase [0]{http://dx.doi.org/}%
\providecommand \selectlanguage [0]{\@gobble}%
\providecommand \bibinfo  [0]{\@secondoftwo}%
\providecommand \bibfield  [0]{\@secondoftwo}%
\providecommand \translation [1]{[#1]}%
\providecommand \BibitemOpen [0]{}%
\providecommand \bibitemStop [0]{}%
\providecommand \bibitemNoStop [0]{.\EOS\space}%
\providecommand \EOS [0]{\spacefactor3000\relax}%
\providecommand \BibitemShut  [1]{\csname bibitem#1\endcsname}%
\let\auto@bib@innerbib\@empty
%</preamble>
\bibitem [{\citenamefont {Nie}(2012)}]{nie2012precipitation}%
  \BibitemOpen
  \bibfield  {author} {\bibinfo {author} {\bibfnamefont {J.-F.}\ \bibnamefont
  {Nie}},\ }\href@noop {} {\bibfield  {journal} {\bibinfo  {journal}
  {Metallurgical and Materials Transactions A}\ }\textbf {\bibinfo {volume}
  {43}},\ \bibinfo {pages} {3891} (\bibinfo {year} {2012})}\BibitemShut
  {NoStop}%
\bibitem [{\citenamefont {Gladman}(1999)}]{gladman1999precipitation}%
  \BibitemOpen
  \bibfield  {author} {\bibinfo {author} {\bibfnamefont {T.}~\bibnamefont
  {Gladman}},\ }\href@noop {} {\bibfield  {journal} {\bibinfo  {journal}
  {Materials Science and Technology}\ }\textbf {\bibinfo {volume} {15}},\
  \bibinfo {pages} {30} (\bibinfo {year} {1999})}\BibitemShut {NoStop}%
\bibitem [{\citenamefont {Martin}(2012)}]{martin2012precipitation}%
  \BibitemOpen
  \bibfield  {author} {\bibinfo {author} {\bibfnamefont {J.~W.}\ \bibnamefont
  {Martin}},\ }\href@noop {} {\emph {\bibinfo {title} {Precipitation Hardening:
  Theory and Applications}}}\ (\bibinfo  {publisher} {Butterworth-Heinemann,
  Oxford},\ \bibinfo {year} {2012})\BibitemShut {NoStop}%
\bibitem [{\citenamefont {Ratke}\ and\ \citenamefont
  {Voorhees}(2013)}]{ratke2013growth}%
  \BibitemOpen
  \bibfield  {author} {\bibinfo {author} {\bibfnamefont {L.}~\bibnamefont
  {Ratke}}\ and\ \bibinfo {author} {\bibfnamefont {P.~W.}\ \bibnamefont
  {Voorhees}},\ }\href@noop {} {\emph {\bibinfo {title} {Growth and Coarsening:
  Ostwald Ripening in Material Processing}}}\ (\bibinfo  {publisher} {Springer
  Science \& Business Media},\ \bibinfo {year} {2013})\BibitemShut {NoStop}%
\bibitem [{\citenamefont {Raabe}\ \emph {et~al.}(2014)\citenamefont {Raabe},
  \citenamefont {Herbig}, \citenamefont {Sandl{\"o}bes}, \citenamefont {Li},
  \citenamefont {Tytko}, \citenamefont {Kuzmina}, \citenamefont {Ponge},\ and\
  \citenamefont {Choi}}]{raabe2014grain}%
  \BibitemOpen
  \bibfield  {author} {\bibinfo {author} {\bibfnamefont {D.}~\bibnamefont
  {Raabe}}, \bibinfo {author} {\bibfnamefont {M.}~\bibnamefont {Herbig}},
  \bibinfo {author} {\bibfnamefont {S.}~\bibnamefont {Sandl{\"o}bes}}, \bibinfo
  {author} {\bibfnamefont {Y.}~\bibnamefont {Li}}, \bibinfo {author}
  {\bibfnamefont {D.}~\bibnamefont {Tytko}}, \bibinfo {author} {\bibfnamefont
  {M.}~\bibnamefont {Kuzmina}}, \bibinfo {author} {\bibfnamefont
  {D.}~\bibnamefont {Ponge}}, \ and\ \bibinfo {author} {\bibfnamefont {P.-P.}\
  \bibnamefont {Choi}},\ }\href@noop {} {\bibfield  {journal} {\bibinfo
  {journal} {Current Opinion in Solid State and Materials Science}\ }\textbf
  {\bibinfo {volume} {18}},\ \bibinfo {pages} {253} (\bibinfo {year}
  {2014})}\BibitemShut {NoStop}%
\bibitem [{\citenamefont {Gleiter}(2000)}]{gleiter2000nanostructured}%
  \BibitemOpen
  \bibfield  {author} {\bibinfo {author} {\bibfnamefont {H.}~\bibnamefont
  {Gleiter}},\ }\href@noop {} {\bibfield  {journal} {\bibinfo  {journal} {Acta
  Materialia}\ }\textbf {\bibinfo {volume} {48}},\ \bibinfo {pages} {1}
  (\bibinfo {year} {2000})}\BibitemShut {NoStop}%
\bibitem [{\citenamefont {Vaithyanathan}\ \emph {et~al.}(2004)\citenamefont
  {Vaithyanathan}, \citenamefont {Wolverton},\ and\ \citenamefont
  {Chen}}]{vaithyanathan2004multiscale}%
  \BibitemOpen
  \bibfield  {author} {\bibinfo {author} {\bibfnamefont {V.}~\bibnamefont
  {Vaithyanathan}}, \bibinfo {author} {\bibfnamefont {C.}~\bibnamefont
  {Wolverton}}, \ and\ \bibinfo {author} {\bibfnamefont {L.}~\bibnamefont
  {Chen}},\ }\href@noop {} {\bibfield  {journal} {\bibinfo  {journal} {Acta
  Materialia}\ }\textbf {\bibinfo {volume} {52}},\ \bibinfo {pages} {2973}
  (\bibinfo {year} {2004})}\BibitemShut {NoStop}%
\bibitem [{\citenamefont {Yang}\ \emph {et~al.}(2016)\citenamefont {Yang},
  \citenamefont {Zhang}, \citenamefont {Shao}, \citenamefont {Wang},
  \citenamefont {Cao}, \citenamefont {Zhang}, \citenamefont {Liu},
  \citenamefont {Chen},\ and\ \citenamefont {Sun}}]{yang2016influence}%
  \BibitemOpen
  \bibfield  {author} {\bibinfo {author} {\bibfnamefont {C.}~\bibnamefont
  {Yang}}, \bibinfo {author} {\bibfnamefont {P.}~\bibnamefont {Zhang}},
  \bibinfo {author} {\bibfnamefont {D.}~\bibnamefont {Shao}}, \bibinfo {author}
  {\bibfnamefont {R.}~\bibnamefont {Wang}}, \bibinfo {author} {\bibfnamefont
  {L.}~\bibnamefont {Cao}}, \bibinfo {author} {\bibfnamefont {J.}~\bibnamefont
  {Zhang}}, \bibinfo {author} {\bibfnamefont {G.}~\bibnamefont {Liu}}, \bibinfo
  {author} {\bibfnamefont {B.}~\bibnamefont {Chen}}, \ and\ \bibinfo {author}
  {\bibfnamefont {J.}~\bibnamefont {Sun}},\ }\href@noop {} {\bibfield
  {journal} {\bibinfo  {journal} {Acta Materialia}\ }\textbf {\bibinfo {volume}
  {119}},\ \bibinfo {pages} {68} (\bibinfo {year} {2016})}\BibitemShut
  {NoStop}%
\bibitem [{\citenamefont {Gibson}\ \emph {et~al.}(2010)\citenamefont {Gibson},
  \citenamefont {Fang}, \citenamefont {Bettles},\ and\ \citenamefont
  {Hutchinson}}]{gibson2010effect}%
  \BibitemOpen
  \bibfield  {author} {\bibinfo {author} {\bibfnamefont {M.}~\bibnamefont
  {Gibson}}, \bibinfo {author} {\bibfnamefont {X.}~\bibnamefont {Fang}},
  \bibinfo {author} {\bibfnamefont {C.}~\bibnamefont {Bettles}}, \ and\
  \bibinfo {author} {\bibfnamefont {C.}~\bibnamefont {Hutchinson}},\
  }\href@noop {} {\bibfield  {journal} {\bibinfo  {journal} {Scripta
  Materialia}\ }\textbf {\bibinfo {volume} {63}},\ \bibinfo {pages} {899}
  (\bibinfo {year} {2010})}\BibitemShut {NoStop}%
\bibitem [{\citenamefont {Jokisaari}\ \emph {et~al.}(2017)\citenamefont
  {Jokisaari}, \citenamefont {Naghavi}, \citenamefont {Wolverton},
  \citenamefont {Voorhees},\ and\ \citenamefont
  {Heinonen}}]{jokisaari2017predicting}%
  \BibitemOpen
  \bibfield  {author} {\bibinfo {author} {\bibfnamefont {A.~M.}\ \bibnamefont
  {Jokisaari}}, \bibinfo {author} {\bibfnamefont {S.~S.}\ \bibnamefont
  {Naghavi}}, \bibinfo {author} {\bibfnamefont {C.}~\bibnamefont {Wolverton}},
  \bibinfo {author} {\bibfnamefont {P.~W.}\ \bibnamefont {Voorhees}}, \ and\
  \bibinfo {author} {\bibfnamefont {O.~G.}\ \bibnamefont {Heinonen}},\
  }\href@noop {} {\bibfield  {journal} {\bibinfo  {journal} {Acta Materialia}\
  }\textbf {\bibinfo {volume} {141}},\ \bibinfo {pages} {273} (\bibinfo {year}
  {2017})}\BibitemShut {NoStop}%
\bibitem [{\citenamefont {Marquis}\ \emph {et~al.}(2003)\citenamefont
  {Marquis}, \citenamefont {Seidman}, \citenamefont {Asta}, \citenamefont
  {Woodward},\ and\ \citenamefont {Ozoli{\c{n}}{\v{s}}}}]{marquis2003mg}%
  \BibitemOpen
  \bibfield  {author} {\bibinfo {author} {\bibfnamefont {E.}~\bibnamefont
  {Marquis}}, \bibinfo {author} {\bibfnamefont {D.~N.}\ \bibnamefont
  {Seidman}}, \bibinfo {author} {\bibfnamefont {M.}~\bibnamefont {Asta}},
  \bibinfo {author} {\bibfnamefont {C.}~\bibnamefont {Woodward}}, \ and\
  \bibinfo {author} {\bibfnamefont {V.}~\bibnamefont {Ozoli{\c{n}}{\v{s}}}},\
  }\href@noop {} {\bibfield  {journal} {\bibinfo  {journal} {Physical Review
  Letters}\ }\textbf {\bibinfo {volume} {91}},\ \bibinfo {pages} {036101}
  (\bibinfo {year} {2003})}\BibitemShut {NoStop}%
\bibitem [{\citenamefont {Marquis}\ and\ \citenamefont
  {Seidman}(2005)}]{marquis2005coarsening}%
  \BibitemOpen
  \bibfield  {author} {\bibinfo {author} {\bibfnamefont {E.~A.}\ \bibnamefont
  {Marquis}}\ and\ \bibinfo {author} {\bibfnamefont {D.~N.}\ \bibnamefont
  {Seidman}},\ }\href@noop {} {\bibfield  {journal} {\bibinfo  {journal} {Acta
  Materialia}\ }\textbf {\bibinfo {volume} {53}},\ \bibinfo {pages} {4259}
  (\bibinfo {year} {2005})}\BibitemShut {NoStop}%
\bibitem [{\citenamefont {Liu}\ \emph {et~al.}(2016)\citenamefont {Liu},
  \citenamefont {Chen},\ and\ \citenamefont {Nie}}]{liu2016interphase}%
  \BibitemOpen
  \bibfield  {author} {\bibinfo {author} {\bibfnamefont {C.}~\bibnamefont
  {Liu}}, \bibinfo {author} {\bibfnamefont {H.}~\bibnamefont {Chen}}, \ and\
  \bibinfo {author} {\bibfnamefont {J.-F.}\ \bibnamefont {Nie}},\ }\href@noop
  {} {\bibfield  {journal} {\bibinfo  {journal} {Scripta Materialia}\ }\textbf
  {\bibinfo {volume} {123}},\ \bibinfo {pages} {5} (\bibinfo {year}
  {2016})}\BibitemShut {NoStop}%
\bibitem [{\citenamefont {Liu}\ \emph {et~al.}(2017)\citenamefont {Liu},
  \citenamefont {Chen},\ and\ \citenamefont {Nie}}]{liu2017zn}%
  \BibitemOpen
  \bibfield  {author} {\bibinfo {author} {\bibfnamefont {C.}~\bibnamefont
  {Liu}}, \bibinfo {author} {\bibfnamefont {H.}~\bibnamefont {Chen}}, \ and\
  \bibinfo {author} {\bibfnamefont {J.-F.}\ \bibnamefont {Nie}},\ }in\
  \href@noop {} {\emph {\bibinfo {booktitle} {Magnesium Technology 2017}}}\
  (\bibinfo  {publisher} {Springer International Publishing},\ \bibinfo {year}
  {2017})\ pp.\ \bibinfo {pages} {53--59}\BibitemShut {NoStop}%
\bibitem [{\citenamefont {Kang}\ \emph {et~al.}(2014)\citenamefont {Kang},
  \citenamefont {Kim}, \citenamefont {Kim},\ and\ \citenamefont
  {Zuo}}]{kang2014determination}%
  \BibitemOpen
  \bibfield  {author} {\bibinfo {author} {\bibfnamefont {S.~J.}\ \bibnamefont
  {Kang}}, \bibinfo {author} {\bibfnamefont {Y.-W.}\ \bibnamefont {Kim}},
  \bibinfo {author} {\bibfnamefont {M.}~\bibnamefont {Kim}}, \ and\ \bibinfo
  {author} {\bibfnamefont {J.-M.}\ \bibnamefont {Zuo}},\ }\href@noop {}
  {\bibfield  {journal} {\bibinfo  {journal} {Acta Materialia}\ }\textbf
  {\bibinfo {volume} {81}},\ \bibinfo {pages} {501} (\bibinfo {year}
  {2014})}\BibitemShut {NoStop}%
\bibitem [{\citenamefont {Marquis}\ \emph {et~al.}(2006)\citenamefont
  {Marquis}, \citenamefont {Seidman}, \citenamefont {Asta},\ and\ \citenamefont
  {Woodward}}]{marquis2006composition}%
  \BibitemOpen
  \bibfield  {author} {\bibinfo {author} {\bibfnamefont {E.~A.}\ \bibnamefont
  {Marquis}}, \bibinfo {author} {\bibfnamefont {D.~N.}\ \bibnamefont
  {Seidman}}, \bibinfo {author} {\bibfnamefont {M.}~\bibnamefont {Asta}}, \
  and\ \bibinfo {author} {\bibfnamefont {C.}~\bibnamefont {Woodward}},\
  }\href@noop {} {\bibfield  {journal} {\bibinfo  {journal} {Acta Materialia}\
  }\textbf {\bibinfo {volume} {54}},\ \bibinfo {pages} {119} (\bibinfo {year}
  {2006})}\BibitemShut {NoStop}%
\bibitem [{\citenamefont {Amouyal}\ \emph {et~al.}(2008)\citenamefont
  {Amouyal}, \citenamefont {Mao},\ and\ \citenamefont
  {Seidman}}]{amouyal2008segregation}%
  \BibitemOpen
  \bibfield  {author} {\bibinfo {author} {\bibfnamefont {Y.}~\bibnamefont
  {Amouyal}}, \bibinfo {author} {\bibfnamefont {Z.}~\bibnamefont {Mao}}, \ and\
  \bibinfo {author} {\bibfnamefont {D.~N.}\ \bibnamefont {Seidman}},\
  }\href@noop {} {\bibfield  {journal} {\bibinfo  {journal} {Applied Physics
  Letters}\ }\textbf {\bibinfo {volume} {93}},\ \bibinfo {pages} {201905}
  (\bibinfo {year} {2008})}\BibitemShut {NoStop}%
\bibitem [{\citenamefont {Biswas}\ \emph {et~al.}(2010)\citenamefont {Biswas},
  \citenamefont {Siegel},\ and\ \citenamefont
  {Seidman}}]{biswas2010simultaneous}%
  \BibitemOpen
  \bibfield  {author} {\bibinfo {author} {\bibfnamefont {A.}~\bibnamefont
  {Biswas}}, \bibinfo {author} {\bibfnamefont {D.~J.}\ \bibnamefont {Siegel}},
  \ and\ \bibinfo {author} {\bibfnamefont {D.~N.}\ \bibnamefont {Seidman}},\
  }\href@noop {} {\bibfield  {journal} {\bibinfo  {journal} {Physical Review
  Letters}\ }\textbf {\bibinfo {volume} {105}},\ \bibinfo {pages} {076102}
  (\bibinfo {year} {2010})}\BibitemShut {NoStop}%
\bibitem [{\citenamefont {Biswas}\ \emph {et~al.}(2011)\citenamefont {Biswas},
  \citenamefont {Siegel}, \citenamefont {Wolverton},\ and\ \citenamefont
  {Seidman}}]{biswas2011precipitates}%
  \BibitemOpen
  \bibfield  {author} {\bibinfo {author} {\bibfnamefont {A.}~\bibnamefont
  {Biswas}}, \bibinfo {author} {\bibfnamefont {D.~J.}\ \bibnamefont {Siegel}},
  \bibinfo {author} {\bibfnamefont {C.}~\bibnamefont {Wolverton}}, \ and\
  \bibinfo {author} {\bibfnamefont {D.~N.}\ \bibnamefont {Seidman}},\
  }\href@noop {} {\bibfield  {journal} {\bibinfo  {journal} {Acta Materialia}\
  }\textbf {\bibinfo {volume} {59}},\ \bibinfo {pages} {6187} (\bibinfo {year}
  {2011})}\BibitemShut {NoStop}%
\bibitem [{\citenamefont {Shin}\ \emph {et~al.}(2017)\citenamefont {Shin},
  \citenamefont {Shyam}, \citenamefont {Lee}, \citenamefont {Yamamoto},\ and\
  \citenamefont {Haynes}}]{shin2017solute}%
  \BibitemOpen
  \bibfield  {author} {\bibinfo {author} {\bibfnamefont {D.}~\bibnamefont
  {Shin}}, \bibinfo {author} {\bibfnamefont {A.}~\bibnamefont {Shyam}},
  \bibinfo {author} {\bibfnamefont {S.}~\bibnamefont {Lee}}, \bibinfo {author}
  {\bibfnamefont {Y.}~\bibnamefont {Yamamoto}}, \ and\ \bibinfo {author}
  {\bibfnamefont {J.~A.}\ \bibnamefont {Haynes}},\ }\href@noop {} {\bibfield
  {journal} {\bibinfo  {journal} {Acta Materialia}\ }\textbf {\bibinfo {volume}
  {141}},\ \bibinfo {pages} {327} (\bibinfo {year} {2017})}\BibitemShut
  {NoStop}%
\bibitem [{\citenamefont {Osamura}\ \emph {et~al.}(1986)\citenamefont
  {Osamura}, \citenamefont {Nakamura}, \citenamefont {Kobayashi}, \citenamefont
  {Hashizume},\ and\ \citenamefont {Sakurai}}]{osamura1986ap}%
  \BibitemOpen
  \bibfield  {author} {\bibinfo {author} {\bibfnamefont {K.}~\bibnamefont
  {Osamura}}, \bibinfo {author} {\bibfnamefont {T.}~\bibnamefont {Nakamura}},
  \bibinfo {author} {\bibfnamefont {A.}~\bibnamefont {Kobayashi}}, \bibinfo
  {author} {\bibfnamefont {T.}~\bibnamefont {Hashizume}}, \ and\ \bibinfo
  {author} {\bibfnamefont {T.}~\bibnamefont {Sakurai}},\ }\href@noop {}
  {\bibfield  {journal} {\bibinfo  {journal} {Acta Metallurgica}\ }\textbf
  {\bibinfo {volume} {34}},\ \bibinfo {pages} {1563} (\bibinfo {year}
  {1986})}\BibitemShut {NoStop}%
\bibitem [{\citenamefont {Rosalie}\ \emph {et~al.}(2014)\citenamefont
  {Rosalie}, \citenamefont {Dwyer},\ and\ \citenamefont
  {Bourgeois}}]{rosalie2014chemical}%
  \BibitemOpen
  \bibfield  {author} {\bibinfo {author} {\bibfnamefont {J.~M.}\ \bibnamefont
  {Rosalie}}, \bibinfo {author} {\bibfnamefont {C.}~\bibnamefont {Dwyer}}, \
  and\ \bibinfo {author} {\bibfnamefont {L.}~\bibnamefont {Bourgeois}},\
  }\href@noop {} {\bibfield  {journal} {\bibinfo  {journal} {Acta Materialia}\
  }\textbf {\bibinfo {volume} {69}},\ \bibinfo {pages} {224} (\bibinfo {year}
  {2014})}\BibitemShut {NoStop}%
\bibitem [{\citenamefont {Zhang}\ \emph {et~al.}(2017)\citenamefont {Zhang},
  \citenamefont {Bourgeois}, \citenamefont {Rosalie},\ and\ \citenamefont
  {Medhekar}}]{zhang2017bi}%
  \BibitemOpen
  \bibfield  {author} {\bibinfo {author} {\bibfnamefont {Z.}~\bibnamefont
  {Zhang}}, \bibinfo {author} {\bibfnamefont {L.}~\bibnamefont {Bourgeois}},
  \bibinfo {author} {\bibfnamefont {J.~M.}\ \bibnamefont {Rosalie}}, \ and\
  \bibinfo {author} {\bibfnamefont {N.~V.}\ \bibnamefont {Medhekar}},\
  }\href@noop {} {\bibfield  {journal} {\bibinfo  {journal} {Acta Materialia}\
  }\textbf {\bibinfo {volume} {132}},\ \bibinfo {pages} {525} (\bibinfo {year}
  {2017})}\BibitemShut {NoStop}%
\bibitem [{\citenamefont {Vaithyanathan}\ \emph {et~al.}(2002)\citenamefont
  {Vaithyanathan}, \citenamefont {Wolverton},\ and\ \citenamefont
  {Chen}}]{vaithyanathan2002multiscale}%
  \BibitemOpen
  \bibfield  {author} {\bibinfo {author} {\bibfnamefont {V.}~\bibnamefont
  {Vaithyanathan}}, \bibinfo {author} {\bibfnamefont {C.}~\bibnamefont
  {Wolverton}}, \ and\ \bibinfo {author} {\bibfnamefont {L.~Q.}\ \bibnamefont
  {Chen}},\ }\href@noop {} {\bibfield  {journal} {\bibinfo  {journal} {Physical
  Review Letters}\ }\textbf {\bibinfo {volume} {88}},\ \bibinfo {pages}
  {125503} (\bibinfo {year} {2002})}\BibitemShut {NoStop}%
\bibitem [{\citenamefont {Thornton}\ \emph {et~al.}(2003)\citenamefont
  {Thornton}, \citenamefont {{\AA}gren},\ and\ \citenamefont
  {Voorhees}}]{thornton2003modelling}%
  \BibitemOpen
  \bibfield  {author} {\bibinfo {author} {\bibfnamefont {K.}~\bibnamefont
  {Thornton}}, \bibinfo {author} {\bibfnamefont {J.}~\bibnamefont {{\AA}gren}},
  \ and\ \bibinfo {author} {\bibfnamefont {P.~W.}\ \bibnamefont {Voorhees}},\
  }\href@noop {} {\bibfield  {journal} {\bibinfo  {journal} {Acta Materialia}\
  }\textbf {\bibinfo {volume} {51}},\ \bibinfo {pages} {5675} (\bibinfo {year}
  {2003})}\BibitemShut {NoStop}%
\bibitem [{\citenamefont {Kim}\ \emph {et~al.}(2017)\citenamefont {Kim},
  \citenamefont {Roy}, \citenamefont {Gururajan}, \citenamefont {Wolverton},\
  and\ \citenamefont {Voorhees}}]{kim2017first}%
  \BibitemOpen
  \bibfield  {author} {\bibinfo {author} {\bibfnamefont {K.}~\bibnamefont
  {Kim}}, \bibinfo {author} {\bibfnamefont {A.}~\bibnamefont {Roy}}, \bibinfo
  {author} {\bibfnamefont {M.}~\bibnamefont {Gururajan}}, \bibinfo {author}
  {\bibfnamefont {C.}~\bibnamefont {Wolverton}}, \ and\ \bibinfo {author}
  {\bibfnamefont {P.}~\bibnamefont {Voorhees}},\ }\href@noop {} {\bibfield
  {journal} {\bibinfo  {journal} {Acta Materialia}\ }\textbf {\bibinfo {volume}
  {140}},\ \bibinfo {pages} {344} (\bibinfo {year} {2017})}\BibitemShut
  {NoStop}%
\bibitem [{\citenamefont {Wynblatt}\ and\ \citenamefont
  {Chatain}(2006)}]{wynblatt2006anisotropy}%
  \BibitemOpen
  \bibfield  {author} {\bibinfo {author} {\bibfnamefont {P.}~\bibnamefont
  {Wynblatt}}\ and\ \bibinfo {author} {\bibfnamefont {D.}~\bibnamefont
  {Chatain}},\ }\href@noop {} {\bibfield  {journal} {\bibinfo  {journal}
  {Metallurgical and Materials Transactions A}\ }\textbf {\bibinfo {volume}
  {37}},\ \bibinfo {pages} {2595} (\bibinfo {year} {2006})}\BibitemShut
  {NoStop}%
\bibitem [{\citenamefont {Wheeler}\ \emph {et~al.}(1992)\citenamefont
  {Wheeler}, \citenamefont {Boettinger},\ and\ \citenamefont
  {McFadden}}]{wheeler1992phase}%
  \BibitemOpen
  \bibfield  {author} {\bibinfo {author} {\bibfnamefont {A.~A.}\ \bibnamefont
  {Wheeler}}, \bibinfo {author} {\bibfnamefont {W.~J.}\ \bibnamefont
  {Boettinger}}, \ and\ \bibinfo {author} {\bibfnamefont {G.~B.}\ \bibnamefont
  {McFadden}},\ }\href@noop {} {\bibfield  {journal} {\bibinfo  {journal}
  {Physical Review A}\ }\textbf {\bibinfo {volume} {45}},\ \bibinfo {pages}
  {7424} (\bibinfo {year} {1992})}\BibitemShut {NoStop}%
\bibitem [{\citenamefont {Kim}\ \emph {et~al.}(1999)\citenamefont {Kim},
  \citenamefont {Kim},\ and\ \citenamefont {Suzuki}}]{kim1999phase}%
  \BibitemOpen
  \bibfield  {author} {\bibinfo {author} {\bibfnamefont {S.~G.}\ \bibnamefont
  {Kim}}, \bibinfo {author} {\bibfnamefont {W.~T.}\ \bibnamefont {Kim}}, \ and\
  \bibinfo {author} {\bibfnamefont {T.}~\bibnamefont {Suzuki}},\ }\href@noop {}
  {\bibfield  {journal} {\bibinfo  {journal} {Physical Review E}\ }\textbf
  {\bibinfo {volume} {60}},\ \bibinfo {pages} {7186} (\bibinfo {year}
  {1999})}\BibitemShut {NoStop}%
\bibitem [{\citenamefont {Kim}\ \emph {et~al.}(2004)\citenamefont {Kim},
  \citenamefont {Kim}, \citenamefont {Suzuki},\ and\ \citenamefont
  {Ode}}]{kim2004phase}%
  \BibitemOpen
  \bibfield  {author} {\bibinfo {author} {\bibfnamefont {S.~G.}\ \bibnamefont
  {Kim}}, \bibinfo {author} {\bibfnamefont {W.~T.}\ \bibnamefont {Kim}},
  \bibinfo {author} {\bibfnamefont {T.}~\bibnamefont {Suzuki}}, \ and\ \bibinfo
  {author} {\bibfnamefont {M.}~\bibnamefont {Ode}},\ }\href@noop {} {\bibfield
  {journal} {\bibinfo  {journal} {Journal of Crystal Growth}\ }\textbf
  {\bibinfo {volume} {261}},\ \bibinfo {pages} {135} (\bibinfo {year}
  {2004})}\BibitemShut {NoStop}%
\bibitem [{\citenamefont {McFadden}\ and\ \citenamefont
  {Wheeler}(2002)}]{mcfadden2002gibbs}%
  \BibitemOpen
  \bibfield  {author} {\bibinfo {author} {\bibfnamefont {G.~B.}\ \bibnamefont
  {McFadden}}\ and\ \bibinfo {author} {\bibfnamefont {A.}~\bibnamefont
  {Wheeler}},\ }in\ \href@noop {} {\emph {\bibinfo {booktitle} {Proceedings of
  the Royal Society of London A: Mathematical, Physical and Engineering
  Sciences}}},\ Vol.\ \bibinfo {volume} {458}\ (\bibinfo {organization} {The
  Royal Society},\ \bibinfo {year} {2002})\ pp.\ \bibinfo {pages}
  {1129--1149}\BibitemShut {NoStop}%
\bibitem [{\citenamefont
  {Umantsev}(2001{\natexlab{a}})}]{umantsev2001continuum}%
  \BibitemOpen
  \bibfield  {author} {\bibinfo {author} {\bibfnamefont {A.}~\bibnamefont
  {Umantsev}},\ }\href@noop {} {\bibfield  {journal} {\bibinfo  {journal}
  {Physical Review B}\ }\textbf {\bibinfo {volume} {64}},\ \bibinfo {pages}
  {075419} (\bibinfo {year} {2001}{\natexlab{a}})}\BibitemShut {NoStop}%
\bibitem [{\citenamefont
  {Umantsev}(2001{\natexlab{b}})}]{umantsev2001coherency}%
  \BibitemOpen
  \bibfield  {author} {\bibinfo {author} {\bibfnamefont {A.}~\bibnamefont
  {Umantsev}},\ }\href@noop {} {\bibfield  {journal} {\bibinfo  {journal}
  {Interface Science}\ }\textbf {\bibinfo {volume} {9}},\ \bibinfo {pages}
  {237} (\bibinfo {year} {2001}{\natexlab{b}})}\BibitemShut {NoStop}%
\bibitem [{\citenamefont {Shower}\ \emph
  {et~al.}(2019{\natexlab{a}})\citenamefont {Shower}, \citenamefont {Morris},
  \citenamefont {Shin}, \citenamefont {Radhakrishnan}, \citenamefont
  {Poplawsky},\ and\ \citenamefont {Shyam}}]{shower2019mechanisms}%
  \BibitemOpen
  \bibfield  {author} {\bibinfo {author} {\bibfnamefont {P.}~\bibnamefont
  {Shower}}, \bibinfo {author} {\bibfnamefont {J.}~\bibnamefont {Morris}},
  \bibinfo {author} {\bibfnamefont {D.}~\bibnamefont {Shin}}, \bibinfo {author}
  {\bibfnamefont {B.}~\bibnamefont {Radhakrishnan}}, \bibinfo {author}
  {\bibfnamefont {J.}~\bibnamefont {Poplawsky}}, \ and\ \bibinfo {author}
  {\bibfnamefont {A.}~\bibnamefont {Shyam}},\ }\href@noop {} {\bibfield
  {journal} {\bibinfo  {journal} {Materialia}\ ,\ \bibinfo {pages} {100335}}
  (\bibinfo {year} {2019}{\natexlab{a}})}\BibitemShut {NoStop}%
\bibitem [{\citenamefont {Shower}\ \emph
  {et~al.}(2019{\natexlab{b}})\citenamefont {Shower}, \citenamefont {Morris},
  \citenamefont {Shin}, \citenamefont {Radhakrishnan}, \citenamefont {Allard},\
  and\ \citenamefont {Shyam}}]{shower2019temperature}%
  \BibitemOpen
  \bibfield  {author} {\bibinfo {author} {\bibfnamefont {P.}~\bibnamefont
  {Shower}}, \bibinfo {author} {\bibfnamefont {J.~R.}\ \bibnamefont {Morris}},
  \bibinfo {author} {\bibfnamefont {D.}~\bibnamefont {Shin}}, \bibinfo {author}
  {\bibfnamefont {B.}~\bibnamefont {Radhakrishnan}}, \bibinfo {author}
  {\bibfnamefont {L.~F.}\ \bibnamefont {Allard}}, \ and\ \bibinfo {author}
  {\bibfnamefont {A.}~\bibnamefont {Shyam}},\ }\href@noop {} {\bibfield
  {journal} {\bibinfo  {journal} {Materialia}\ }\textbf {\bibinfo {volume}
  {5}},\ \bibinfo {pages} {100185} (\bibinfo {year}
  {2019}{\natexlab{b}})}\BibitemShut {NoStop}%
\bibitem [{\citenamefont {Hu}\ \emph {et~al.}(2007)\citenamefont {Hu},
  \citenamefont {Murray}, \citenamefont {Weiland}, \citenamefont {Liu},\ and\
  \citenamefont {Chen}}]{hu2007thermodynamic}%
  \BibitemOpen
  \bibfield  {author} {\bibinfo {author} {\bibfnamefont {S.}~\bibnamefont
  {Hu}}, \bibinfo {author} {\bibfnamefont {J.}~\bibnamefont {Murray}}, \bibinfo
  {author} {\bibfnamefont {H.}~\bibnamefont {Weiland}}, \bibinfo {author}
  {\bibfnamefont {Z.}~\bibnamefont {Liu}}, \ and\ \bibinfo {author}
  {\bibfnamefont {L.}~\bibnamefont {Chen}},\ }\href@noop {} {\bibfield
  {journal} {\bibinfo  {journal} {Calphad}\ }\textbf {\bibinfo {volume} {31}},\
  \bibinfo {pages} {303} (\bibinfo {year} {2007})}\BibitemShut {NoStop}%
\bibitem [{\citenamefont {Provatas}\ and\ \citenamefont
  {Elder}(2011)}]{provatas2011phase}%
  \BibitemOpen
  \bibfield  {author} {\bibinfo {author} {\bibfnamefont {N.}~\bibnamefont
  {Provatas}}\ and\ \bibinfo {author} {\bibfnamefont {K.}~\bibnamefont
  {Elder}},\ }\href@noop {} {\emph {\bibinfo {title} {Phase-field Methods in
  Materials Science and Engineering}}}\ (\bibinfo  {publisher} {John Wiley \&
  Sons},\ \bibinfo {year} {2011})\BibitemShut {NoStop}%
\bibitem [{\citenamefont {Dregia}\ and\ \citenamefont
  {Wynblatt}(1991)}]{dregia1991equilibrium}%
  \BibitemOpen
  \bibfield  {author} {\bibinfo {author} {\bibfnamefont {S.}~\bibnamefont
  {Dregia}}\ and\ \bibinfo {author} {\bibfnamefont {P.}~\bibnamefont
  {Wynblatt}},\ }\href@noop {} {\bibfield  {journal} {\bibinfo  {journal} {Acta
  Metallurgica et Materialia}\ }\textbf {\bibinfo {volume} {39}},\ \bibinfo
  {pages} {771} (\bibinfo {year} {1991})}\BibitemShut {NoStop}%
\bibitem [{\citenamefont {Huang}\ \emph {et~al.}(1999)\citenamefont {Huang},
  \citenamefont {De~La~Cruz},\ and\ \citenamefont
  {Voorhees}}]{huang1999interfacial}%
  \BibitemOpen
  \bibfield  {author} {\bibinfo {author} {\bibfnamefont {C.}~\bibnamefont
  {Huang}}, \bibinfo {author} {\bibfnamefont {M.~O.}\ \bibnamefont
  {De~La~Cruz}}, \ and\ \bibinfo {author} {\bibfnamefont {P.~W.}\ \bibnamefont
  {Voorhees}},\ }\href@noop {} {\bibfield  {journal} {\bibinfo  {journal} {Acta
  Materialia}\ }\textbf {\bibinfo {volume} {47}},\ \bibinfo {pages} {4449}
  (\bibinfo {year} {1999})}\BibitemShut {NoStop}%
\bibitem [{\citenamefont {Abdeljawad}\ and\ \citenamefont
  {Foiles}(2015)}]{abdeljawad2015stabilization}%
  \BibitemOpen
  \bibfield  {author} {\bibinfo {author} {\bibfnamefont {F.}~\bibnamefont
  {Abdeljawad}}\ and\ \bibinfo {author} {\bibfnamefont {S.~M.}\ \bibnamefont
  {Foiles}},\ }\href@noop {} {\bibfield  {journal} {\bibinfo  {journal} {Acta
  Materialia}\ }\textbf {\bibinfo {volume} {101}},\ \bibinfo {pages} {159}
  (\bibinfo {year} {2015})}\BibitemShut {NoStop}%
\bibitem [{\citenamefont {Abdeljawad}\ \emph {et~al.}(2017)\citenamefont
  {Abdeljawad}, \citenamefont {Lu}, \citenamefont {Argibay}, \citenamefont
  {Clark}, \citenamefont {Boyce},\ and\ \citenamefont
  {Foiles}}]{abdeljawad2017grain}%
  \BibitemOpen
  \bibfield  {author} {\bibinfo {author} {\bibfnamefont {F.}~\bibnamefont
  {Abdeljawad}}, \bibinfo {author} {\bibfnamefont {P.}~\bibnamefont {Lu}},
  \bibinfo {author} {\bibfnamefont {N.}~\bibnamefont {Argibay}}, \bibinfo
  {author} {\bibfnamefont {B.~G.}\ \bibnamefont {Clark}}, \bibinfo {author}
  {\bibfnamefont {B.~L.}\ \bibnamefont {Boyce}}, \ and\ \bibinfo {author}
  {\bibfnamefont {S.~M.}\ \bibnamefont {Foiles}},\ }\href@noop {} {\bibfield
  {journal} {\bibinfo  {journal} {Acta Materialia}\ }\textbf {\bibinfo {volume}
  {126}},\ \bibinfo {pages} {528} (\bibinfo {year} {2017})}\BibitemShut
  {NoStop}%
\bibitem [{\citenamefont {Kim}\ \emph {et~al.}(2016)\citenamefont {Kim},
  \citenamefont {Lee},\ and\ \citenamefont {Lee}}]{kim2016GBsegregation}%
  \BibitemOpen
  \bibfield  {author} {\bibinfo {author} {\bibfnamefont {S.~G.}\ \bibnamefont
  {Kim}}, \bibinfo {author} {\bibfnamefont {J.~S.}\ \bibnamefont {Lee}}, \ and\
  \bibinfo {author} {\bibfnamefont {B.-J.}\ \bibnamefont {Lee}},\ }\href@noop
  {} {\bibfield  {journal} {\bibinfo  {journal} {Acta Materialia}\ }\textbf
  {\bibinfo {volume} {112}},\ \bibinfo {pages} {150} (\bibinfo {year}
  {2016})}\BibitemShut {NoStop}%
\bibitem [{\citenamefont {Kaplan}\ \emph {et~al.}(2013)\citenamefont {Kaplan},
  \citenamefont {Chatain}, \citenamefont {Wynblatt},\ and\ \citenamefont
  {Carter}}]{kaplan2013review}%
  \BibitemOpen
  \bibfield  {author} {\bibinfo {author} {\bibfnamefont {W.~D.}\ \bibnamefont
  {Kaplan}}, \bibinfo {author} {\bibfnamefont {D.}~\bibnamefont {Chatain}},
  \bibinfo {author} {\bibfnamefont {P.}~\bibnamefont {Wynblatt}}, \ and\
  \bibinfo {author} {\bibfnamefont {W.~C.}\ \bibnamefont {Carter}},\
  }\href@noop {} {\bibfield  {journal} {\bibinfo  {journal} {Journal of
  Materials Science}\ }\textbf {\bibinfo {volume} {48}},\ \bibinfo {pages}
  {5681} (\bibinfo {year} {2013})}\BibitemShut {NoStop}%
\bibitem [{\citenamefont {Frolov}\ and\ \citenamefont
  {Mishin}(2015)}]{frolov2015phases}%
  \BibitemOpen
  \bibfield  {author} {\bibinfo {author} {\bibfnamefont {T.}~\bibnamefont
  {Frolov}}\ and\ \bibinfo {author} {\bibfnamefont {Y.}~\bibnamefont
  {Mishin}},\ }\href@noop {} {\bibfield  {journal} {\bibinfo  {journal} {The
  Journal of Chemical Physics}\ }\textbf {\bibinfo {volume} {143}},\ \bibinfo
  {pages} {044706} (\bibinfo {year} {2015})}\BibitemShut {NoStop}%
\bibitem [{\citenamefont {Howe}(1997)}]{howe1997interfaces}%
  \BibitemOpen
  \bibfield  {author} {\bibinfo {author} {\bibfnamefont {J.~M.}\ \bibnamefont
  {Howe}},\ }\href@noop {} {\emph {\bibinfo {title} {Interfaces in
  Materials}}}\ (\bibinfo  {publisher} {John Wiley \& Sons},\ \bibinfo {year}
  {1997})\BibitemShut {NoStop}%
\bibitem [{\citenamefont {Kadambi}\ and\ \citenamefont
  {Patala}(2017)}]{kadambi2017thermodynamic}%
  \BibitemOpen
  \bibfield  {author} {\bibinfo {author} {\bibfnamefont {S.~B.}\ \bibnamefont
  {Kadambi}}\ and\ \bibinfo {author} {\bibfnamefont {S.}~\bibnamefont
  {Patala}},\ }\href@noop {} {\bibfield  {journal} {\bibinfo  {journal}
  {Physical Review Materials}\ }\textbf {\bibinfo {volume} {1}},\ \bibinfo
  {pages} {043604} (\bibinfo {year} {2017})}\BibitemShut {NoStop}%
\bibitem [{\citenamefont {Balluffi}\ \emph {et~al.}(2005)\citenamefont
  {Balluffi}, \citenamefont {Allen},\ and\ \citenamefont
  {Carter}}]{balluffi2005kinetics}%
  \BibitemOpen
  \bibfield  {author} {\bibinfo {author} {\bibfnamefont {R.~W.}\ \bibnamefont
  {Balluffi}}, \bibinfo {author} {\bibfnamefont {S.}~\bibnamefont {Allen}}, \
  and\ \bibinfo {author} {\bibfnamefont {W.~C.}\ \bibnamefont {Carter}},\
  }\href@noop {} {\emph {\bibinfo {title} {Kinetics of Materials}}}\ (\bibinfo
  {publisher} {John Wiley \& Sons, Hoboken, NJ},\ \bibinfo {year}
  {2005})\BibitemShut {NoStop}%
\bibitem [{\citenamefont {Hillert}(1975)}]{hillert1975lectures}%
  \BibitemOpen
  \bibfield  {author} {\bibinfo {author} {\bibfnamefont {M.}~\bibnamefont
  {Hillert}},\ }in\ \href@noop {} {\emph {\bibinfo {booktitle} {Lectures on the
  Theory of Phase Transformations}}},\ \bibinfo {editor} {edited by\ \bibinfo
  {editor} {\bibfnamefont {H.~I.}\ \bibnamefont {Aaronson}}}\ (\bibinfo
  {publisher} {AIME},\ \bibinfo {address} {New York},\ \bibinfo {year} {1975})\
  Chap.~\bibinfo {chapter} {1}, pp.\ \bibinfo {pages} {1--50}\BibitemShut
  {NoStop}%
\bibitem [{\citenamefont {Wynblatt}\ and\ \citenamefont
  {Wu}(1979)}]{wynblatt1979interfacial}%
  \BibitemOpen
  \bibfield  {author} {\bibinfo {author} {\bibfnamefont {P.}~\bibnamefont
  {Wynblatt}}\ and\ \bibinfo {author} {\bibfnamefont {R.~C.}\ \bibnamefont
  {Wu}},\ }in\ \href@noop {} {\emph {\bibinfo {booktitle} {Interfacial
  Segregation}}},\ \bibinfo {editor} {edited by\ \bibinfo {editor}
  {\bibfnamefont {W.~C.}\ \bibnamefont {Johnson}}\ and\ \bibinfo {editor}
  {\bibfnamefont {J.~M.}\ \bibnamefont {Blakely}}}\ (\bibinfo  {publisher}
  {ASM},\ \bibinfo {address} {Metals Park, OH},\ \bibinfo {year} {1979})\ pp.\
  \bibinfo {pages} {115--136}\BibitemShut {NoStop}%
\bibitem [{\citenamefont {Sutton}\ and\ \citenamefont
  {Balluffi}(1995)}]{sutton1995interfaces}%
  \BibitemOpen
  \bibfield  {author} {\bibinfo {author} {\bibfnamefont {A.~P.}\ \bibnamefont
  {Sutton}}\ and\ \bibinfo {author} {\bibfnamefont {R.~W.}\ \bibnamefont
  {Balluffi}},\ }\href@noop {} {\emph {\bibinfo {title} {Interfaces in
  Crystalline Materials}}}\ (\bibinfo  {publisher} {Clarendon Press, Oxford},\
  \bibinfo {year} {1995})\BibitemShut {NoStop}%
\bibitem [{\citenamefont {Johnson}(1979)}]{johnson1979interfacial}%
  \BibitemOpen
  \bibfield  {author} {\bibinfo {author} {\bibfnamefont {W.~C.}\ \bibnamefont
  {Johnson}},\ }in\ \href@noop {} {\emph {\bibinfo {booktitle} {Interfacial
  Segregation}}},\ \bibinfo {editor} {edited by\ \bibinfo {editor}
  {\bibfnamefont {W.~C.}\ \bibnamefont {Johnson}}\ and\ \bibinfo {editor}
  {\bibfnamefont {J.~M.}\ \bibnamefont {Blakely}}}\ (\bibinfo  {publisher}
  {ASM},\ \bibinfo {address} {Metals Park, OH},\ \bibinfo {year} {1979})\ pp.\
  \bibinfo {pages} {351--379}\BibitemShut {NoStop}%
\bibitem [{\citenamefont {Cahn}(1979)}]{cahn1979interfacial}%
  \BibitemOpen
  \bibfield  {author} {\bibinfo {author} {\bibfnamefont {J.~W.}\ \bibnamefont
  {Cahn}},\ }in\ \href@noop {} {\emph {\bibinfo {booktitle} {Interfacial
  Segregation}}},\ \bibinfo {editor} {edited by\ \bibinfo {editor}
  {\bibfnamefont {W.~C.}\ \bibnamefont {Johnson}}\ and\ \bibinfo {editor}
  {\bibfnamefont {J.~M.}\ \bibnamefont {Blakely}}}\ (\bibinfo  {publisher}
  {ASM},\ \bibinfo {address} {Metals Park, OH},\ \bibinfo {year} {1979})\ pp.\
  \bibinfo {pages} {3--23}\BibitemShut {NoStop}%
\bibitem [{\citenamefont {Gaskell}\ and\ \citenamefont
  {Laughlin}(2017)}]{gaskell2017introduction}%
  \BibitemOpen
  \bibfield  {author} {\bibinfo {author} {\bibfnamefont {D.~R.}\ \bibnamefont
  {Gaskell}}\ and\ \bibinfo {author} {\bibfnamefont {D.~E.}\ \bibnamefont
  {Laughlin}},\ }\href@noop {} {\emph {\bibinfo {title} {Introduction to the
  Thermodynamics of Materials}}}\ (\bibinfo  {publisher} {CRC press, Boca
  Raton},\ \bibinfo {year} {2017})\BibitemShut {NoStop}%
\bibitem [{\citenamefont {Hillert}(2007)}]{hillert2007phase}%
  \BibitemOpen
  \bibfield  {author} {\bibinfo {author} {\bibfnamefont {M.}~\bibnamefont
  {Hillert}},\ }\href@noop {} {\emph {\bibinfo {title} {Phase Equilibria, Phase
  Diagrams and Phase Transformations: their Thermodynamic Basis}}}\ (\bibinfo
  {publisher} {Cambridge University Press, Oxford},\ \bibinfo {year}
  {2007})\BibitemShut {NoStop}%
\bibitem [{\citenamefont {Lejcek}(2010)}]{lejcek2010grain}%
  \BibitemOpen
  \bibfield  {author} {\bibinfo {author} {\bibfnamefont {P.}~\bibnamefont
  {Lejcek}},\ }\href@noop {} {\emph {\bibinfo {title} {Grain Boundary
  Segregation in Metals}}},\ Vol.\ \bibinfo {volume} {136}\ (\bibinfo
  {publisher} {Springer Science \& Business Media},\ \bibinfo {year}
  {2010})\BibitemShut {NoStop}%
\bibitem [{\citenamefont {Cha}\ \emph {et~al.}(2002)\citenamefont {Cha},
  \citenamefont {Kim}, \citenamefont {Yeon},\ and\ \citenamefont
  {Yoon}}]{cha2002phase}%
  \BibitemOpen
  \bibfield  {author} {\bibinfo {author} {\bibfnamefont {P.-R.}\ \bibnamefont
  {Cha}}, \bibinfo {author} {\bibfnamefont {S.~G.}\ \bibnamefont {Kim}},
  \bibinfo {author} {\bibfnamefont {D.-H.}\ \bibnamefont {Yeon}}, \ and\
  \bibinfo {author} {\bibfnamefont {J.-K.}\ \bibnamefont {Yoon}},\ }\href@noop
  {} {\bibfield  {journal} {\bibinfo  {journal} {Acta materialia}\ }\textbf
  {\bibinfo {volume} {50}},\ \bibinfo {pages} {3817} (\bibinfo {year}
  {2002})}\BibitemShut {NoStop}%
\bibitem [{\citenamefont {Guggenheim}(1985)}]{guggenheim1985thermodynamics}%
  \BibitemOpen
  \bibfield  {author} {\bibinfo {author} {\bibfnamefont {E.~A.}\ \bibnamefont
  {Guggenheim}},\ }\href@noop {} {\emph {\bibinfo {title} {Thermodynamics. An
  Advanced Treatment for Chemists and Physicists}}}\ (\bibinfo  {publisher}
  {North-Holland, Amsterdam},\ \bibinfo {year} {1985})\BibitemShut {NoStop}%
\bibitem [{\citenamefont {Fowler}\ and\ \citenamefont
  {Guggenheim}(1939)}]{fowler1941statistical}%
  \BibitemOpen
  \bibfield  {author} {\bibinfo {author} {\bibfnamefont {R.~H.}\ \bibnamefont
  {Fowler}}\ and\ \bibinfo {author} {\bibfnamefont {E.~A.}\ \bibnamefont
  {Guggenheim}},\ }\href@noop {} {\emph {\bibinfo {title} {Statistical
  Thermodynamics}}}\ (\bibinfo  {publisher} {MacMillan, New York},\ \bibinfo
  {year} {1939})\BibitemShut {NoStop}%
\bibitem [{\citenamefont {McLean}(1957)}]{mcleangrain}%
  \BibitemOpen
  \bibfield  {author} {\bibinfo {author} {\bibfnamefont {D.}~\bibnamefont
  {McLean}},\ }\href@noop {} {\emph {\bibinfo {title} {Grain Boundaries in
  Metals}}}\ (\bibinfo  {publisher} {Oxford University Press, Oxford},\
  \bibinfo {year} {1957})\BibitemShut {NoStop}%
\bibitem [{\citenamefont {da~Silva}\ \emph {et~al.}(2019)\citenamefont
  {da~Silva}, \citenamefont {Kamachali}, \citenamefont {Ponge}, \citenamefont
  {Gault}, \citenamefont {Neugebauer},\ and\ \citenamefont
  {Raabe}}]{da2019thermodynamics}%
  \BibitemOpen
  \bibfield  {author} {\bibinfo {author} {\bibfnamefont {A.~K.}\ \bibnamefont
  {da~Silva}}, \bibinfo {author} {\bibfnamefont {R.~D.}\ \bibnamefont
  {Kamachali}}, \bibinfo {author} {\bibfnamefont {D.}~\bibnamefont {Ponge}},
  \bibinfo {author} {\bibfnamefont {B.}~\bibnamefont {Gault}}, \bibinfo
  {author} {\bibfnamefont {J.}~\bibnamefont {Neugebauer}}, \ and\ \bibinfo
  {author} {\bibfnamefont {D.}~\bibnamefont {Raabe}},\ }\href@noop {}
  {\bibfield  {journal} {\bibinfo  {journal} {Acta Materialia}\ }\textbf
  {\bibinfo {volume} {168}},\ \bibinfo {pages} {109} (\bibinfo {year}
  {2019})}\BibitemShut {NoStop}%
\bibitem [{\citenamefont {Ardell}\ and\ \citenamefont
  {Nicholson}(1966)}]{ardell1966coarsening}%
  \BibitemOpen
  \bibfield  {author} {\bibinfo {author} {\bibfnamefont {A.}~\bibnamefont
  {Ardell}}\ and\ \bibinfo {author} {\bibfnamefont {R.}~\bibnamefont
  {Nicholson}},\ }\href@noop {} {\bibfield  {journal} {\bibinfo  {journal}
  {Journal of Physics and Chemistry of Solids}\ }\textbf {\bibinfo {volume}
  {27}},\ \bibinfo {pages} {1793} (\bibinfo {year} {1966})}\BibitemShut
  {NoStop}%
\bibitem [{\citenamefont {Hyland}(1992)}]{hyland1992homogeneous}%
  \BibitemOpen
  \bibfield  {author} {\bibinfo {author} {\bibfnamefont {R.}~\bibnamefont
  {Hyland}},\ }\href@noop {} {\bibfield  {journal} {\bibinfo  {journal}
  {Metallurgical Transactions A}\ }\textbf {\bibinfo {volume} {23}},\ \bibinfo
  {pages} {1947} (\bibinfo {year} {1992})}\BibitemShut {NoStop}%
\bibitem [{\citenamefont {Krakauer}\ and\ \citenamefont
  {Seidman}(1993)}]{krakauer1993absolute}%
  \BibitemOpen
  \bibfield  {author} {\bibinfo {author} {\bibfnamefont {B.~W.}\ \bibnamefont
  {Krakauer}}\ and\ \bibinfo {author} {\bibfnamefont {D.~N.}\ \bibnamefont
  {Seidman}},\ }\href@noop {} {\bibfield  {journal} {\bibinfo  {journal}
  {Physical Review B}\ }\textbf {\bibinfo {volume} {48}},\ \bibinfo {pages}
  {6724} (\bibinfo {year} {1993})}\BibitemShut {NoStop}%
\bibitem [{\citenamefont {Wheeler}\ \emph {et~al.}(1993)\citenamefont
  {Wheeler}, \citenamefont {Boettinger},\ and\ \citenamefont
  {McFadden}}]{wheeler1993phase}%
  \BibitemOpen
  \bibfield  {author} {\bibinfo {author} {\bibfnamefont {A.~A.}\ \bibnamefont
  {Wheeler}}, \bibinfo {author} {\bibfnamefont {W.~J.}\ \bibnamefont
  {Boettinger}}, \ and\ \bibinfo {author} {\bibfnamefont {G.~B.}\ \bibnamefont
  {McFadden}},\ }\href@noop {} {\bibfield  {journal} {\bibinfo  {journal}
  {Physical Review E}\ }\textbf {\bibinfo {volume} {47}},\ \bibinfo {pages}
  {1893} (\bibinfo {year} {1993})}\BibitemShut {NoStop}%
\bibitem [{\citenamefont {Lupis}(1983)}]{lupis1983chemical}%
  \BibitemOpen
  \bibfield  {author} {\bibinfo {author} {\bibfnamefont {C.~H.}\ \bibnamefont
  {Lupis}},\ }\href@noop {} {\emph {\bibinfo {title} {Chemical Thermodynamics
  of Materials}}}\ (\bibinfo  {publisher} {North-Holland, New York},\ \bibinfo
  {year} {1983})\BibitemShut {NoStop}%
\end{thebibliography}%

% \bibliographystyle{apsrev4-1} % Tell bibtex which bibliography style to use
% \bibliography{main} % Tell bibtex which .bib file to use (this one is some example file in TexLive's file tree)

% \end{bibunit}

% \clearpage
% \input{SupportingInfo.tex}

\end{document}